\documentclass[%
 reprint,
showpacs,
showkeys,
 amsmath,amssymb,
 aps,
]{revtex4-1}

\usepackage{graphicx}
\usepackage{dcolumn}
\usepackage{bm}

\begin{document}


\title{Nuclear neutrino energy spectra in high temperature astrophysical environments}

\author{G. Wendell Misch}
\affiliation{Department of Physics and Astronomy, Shanghai Jiao Tong University, Shanghai 200240, China}
\affiliation{Collaborative Innovation Center of IFSA (CICIFSA), Shanghai Jiao Tong University, Shanghai 200240, China}

\author{George M. Fuller}
\affiliation{Department of Physics, University of California, San Diego, La Jolla, California 92093, USA}

\date{\today}

\begin{abstract}
Astrophysical environments that reach temperatures greater than $\sim 100$ keV can have significant neutrino energy loss via both plasma processes and  nuclear weak interactions.  We find that nuclear processes likely produce the highest-energy neutrinos.  Among the important weak nuclear interactions are both charged current channels (electron capture/emission and positron capture/emission) and neutral current channels (de-excitation of nuclei via neutrino pair emission).  We show that in order to make a realistic prediction of the nuclear neutrino spectrum, one must take nuclear structure into account; in some cases, the most important transitions may involve excited states, possibly in both parent and daughter nuclei.  We find that the standard technique of producing a neutrino energy spectrum by using a single transition with a Q-value and matrix element chosen to fit published neutrino production rates and energy losses will not accurately capture important spectral features.

\end{abstract}

\maketitle


\section{Introduction}
In this paper, we calculate energy spectra for neutrinos produced in nuclear weak interaction processes that occur in pre-collapse massive stars.  A key motivation for this work is the possibility of detecting a neutrino signal from a massive star, perhaps even months before collapse \cite{omk:2004a,omk:2004b,oh:2010,asakura-etal:2016}.  Patton \& Lunardini (hereafter P\&L) \cite{pl:2015} have studied the neutrino emissivity physics in this problem and the associated prospects for detection.  In this paper, we build on the work of P\&L, but we concentrate on the nuclear physics which determines important aspects of the neutrino energy spectra, especially at high neutrino energy.  Higher neutrino energies are, of course, key to detection.  Our nuclear structure considerations and our shell model calculations allow us to illuminate features of specific sd-shell nuclei which are likely to be key contributors to the high-energy end of the expected neutrino spectrum.

Beginning with core carbon burning, neutrino production dominates the energy loss of massive stars.  Depending on the mass of the star and its stage of burning, these neutrinos can be produced through electron-positron pair annihilation, the photo process (wherein a photon interacts with an electron and produces a neutrino pair), electron neutrino pair bremsstrahlung, electron capture, and other processes.  In low mass stars and in massive stars prior to core collapse, the neutrinos stream unimpeded through stellar material, removing entropy from the core and greatly accelerating the evolution of the star \cite{whw:2002}.

\begin{figure}[here]
\includegraphics[scale=.6]{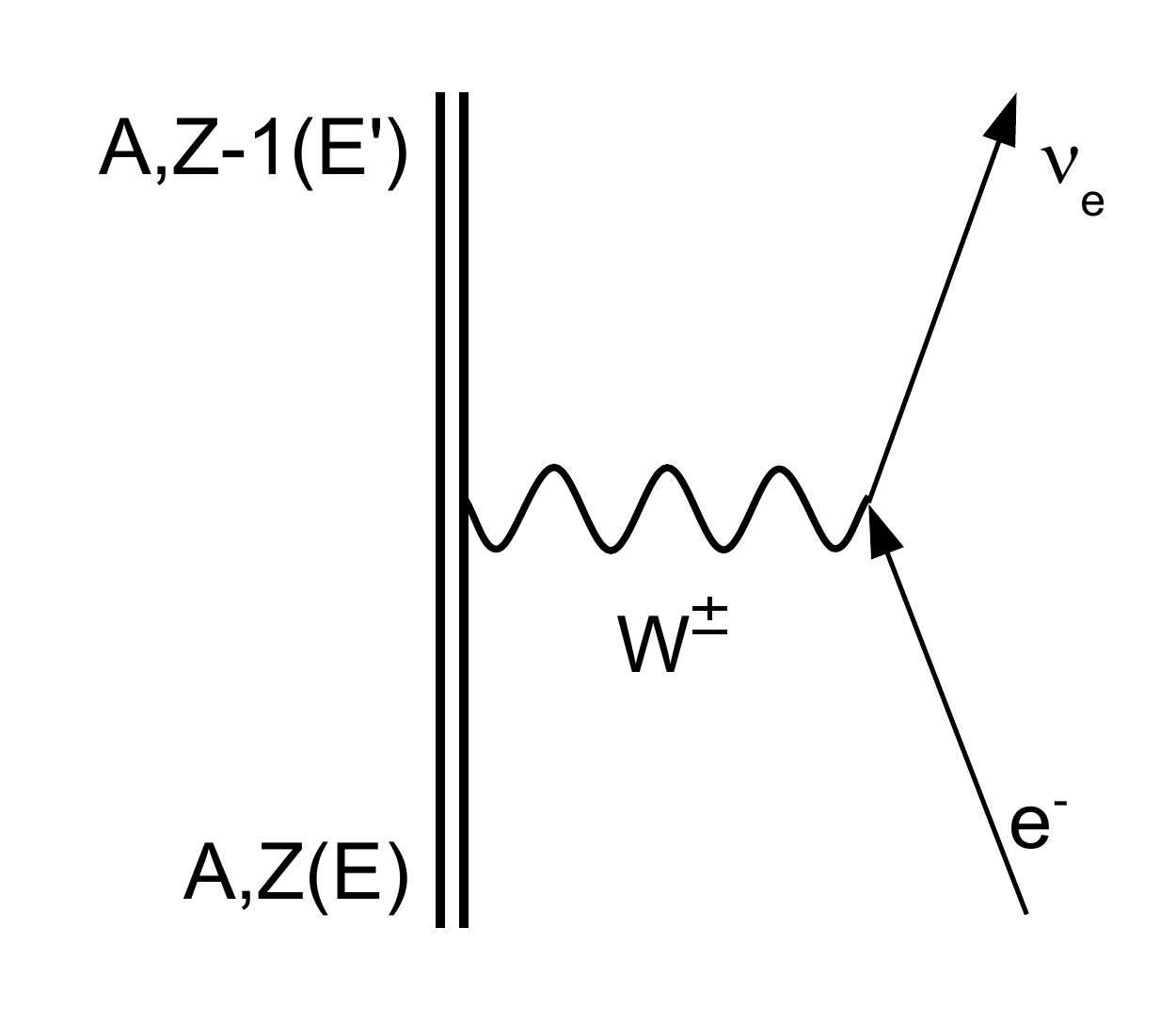}
\caption{Electron capture on a nucleus of mass number A, proton number Z, and excitation energy E, producing a nucleus of mass number A, proton number Z-1, and excitation energy E$'$.  The electron and neutrino may be exchanged in this diagram for their antiparticles, yielding a final nucleus with proton number Z+1.}
\label{fig:feyn_cap}
\end{figure}

\begin{figure}[here]
\includegraphics[scale=.6]{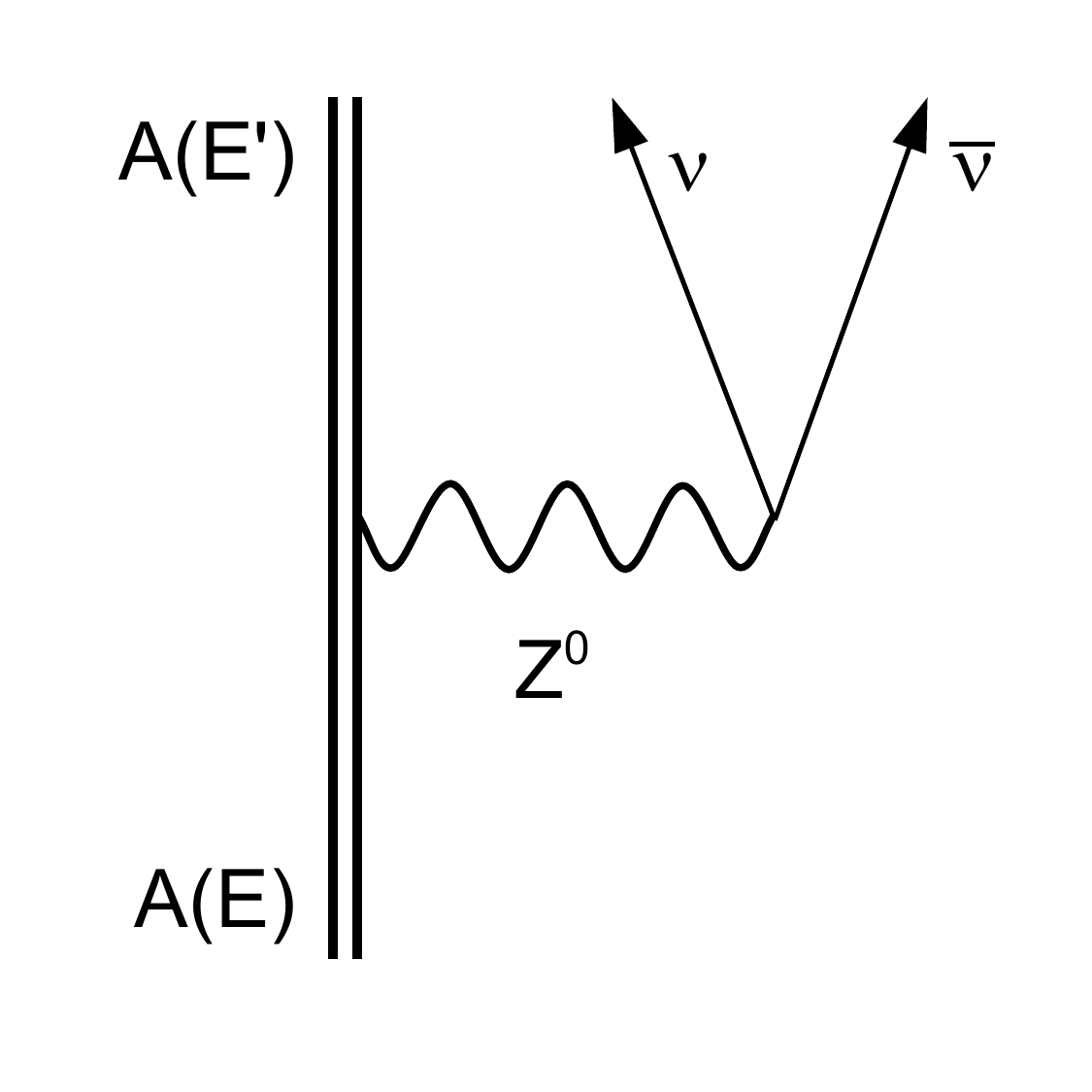}
\caption{Neutral current neutrino pair emission from a nucleus of mass number A with initial excitation energy E and final excitation E$'$.}
\label{fig:feyn_nc_decay}
\end{figure}

\begin{figure}[here]
\includegraphics[scale=.6]{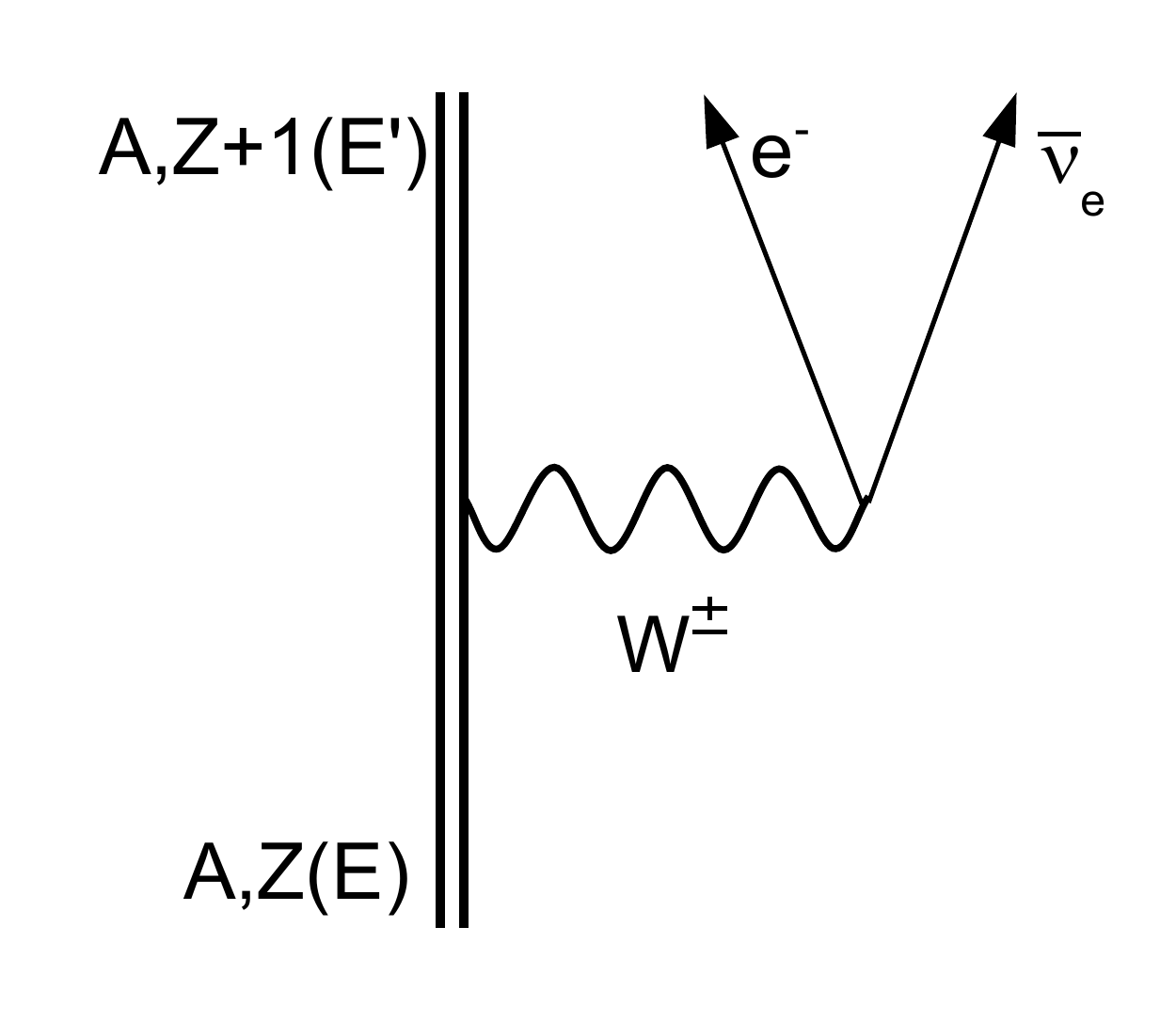}
\caption{Electron decay from a nucleus of mass number A, proton number Z, and excitation energy E to a nucleus of mass number A, proton number Z+1, and excitation energy E$'$.  The electron and antineutrino may be exchanged in this diagram for their antiparticles, yielding a final nucleus with proton number Z-1.}
\label{fig:feyn_dec}
\end{figure}

In the final stages before collapse of a massive star, the core is hot and dense, but the entropy is low \cite{bbal:1979}.  The temperature is $\sim 0.5$ MeV, but the electron Fermi energy can be $\sim 5$ MeV, implying very electron-degenerate conditions \cite{arnett:1977,bw:1982,bw:1985,bmd:2003,bm:2007,sjfk:2008,bbol:2011,ajs:2007,liebendorfer-etal:2008,liebendorfer-etal:2009,hjm:2010,hix-etal:2010,bdm:2012}.  The electron degeneracy relatively suppresses neutrino production processes with electrons in the final state, processes with intermediate electron loops, and electron-positron annihilation.  At the same time, the high Fermi energy relatively enhances electron capture (figure \ref{fig:feyn_cap}), while the high temperature gives a population of excited nucleons that can de-excite by emission of a neutrino pair (figure \ref{fig:feyn_nc_decay}) \cite{btz:1974,fm:1991,mbf:2013}.  In many cases, excited nuclei can also more readily decay by electron or positron emission \cite{ffn:1982a}, which is always accompanied by an anti-neutrino or neutrino, respectively (figure \ref{fig:feyn_dec}).

High temperatures allow the nuclei to access excited parent states which may have large Q-values and large weak interaction matrix elements for charged current transitions.  Large Q-values imply larger phase space factors for weak interactions, but against this, Boltzmann population factors for these highly excited initial states can be small.  However, ameliorating the effect of small Boltzmann factors is the near-exponential increase in nuclear level densities with increasing excitation energies.  In the end, the balance between all these factors must be evaluated on a case-by-case basis for individual nuclei and particular thermodynamic conditions in the star.

This situation has a profound effect on the neutrino spectrum, as energetic electrons can capture onto excited parent nuclei, which might have a \emph{less} excited final state in the daughter nucleus; this results in an unusually high energy neutrino.

Furthermore, these excited nuclei may directly produce neutrino pairs.  When excited nuclei de-excite, the usual channel is gamma ray emission; however, they may also emit a virtual Z$^0$ boson that decays into a neutrino anti-neutrino pair, shown schematically in figure \ref{fig:feyn_nc_decay}.  In fact, this can be the dominant source of neutrino pairs in a collapsing stellar core \cite{gershtein-etal:1975,km:1979,fm:1991,mbf:2013,flm:2013}.  If the nucleus de-excite from a highly excited state, it can produce an energetic neutrino pair of any flavor, and these neutrinos can make a substantial contribution at the high energy end of the neutrino spectrum.

\begin{figure}[here]
\includegraphics[scale=.6]{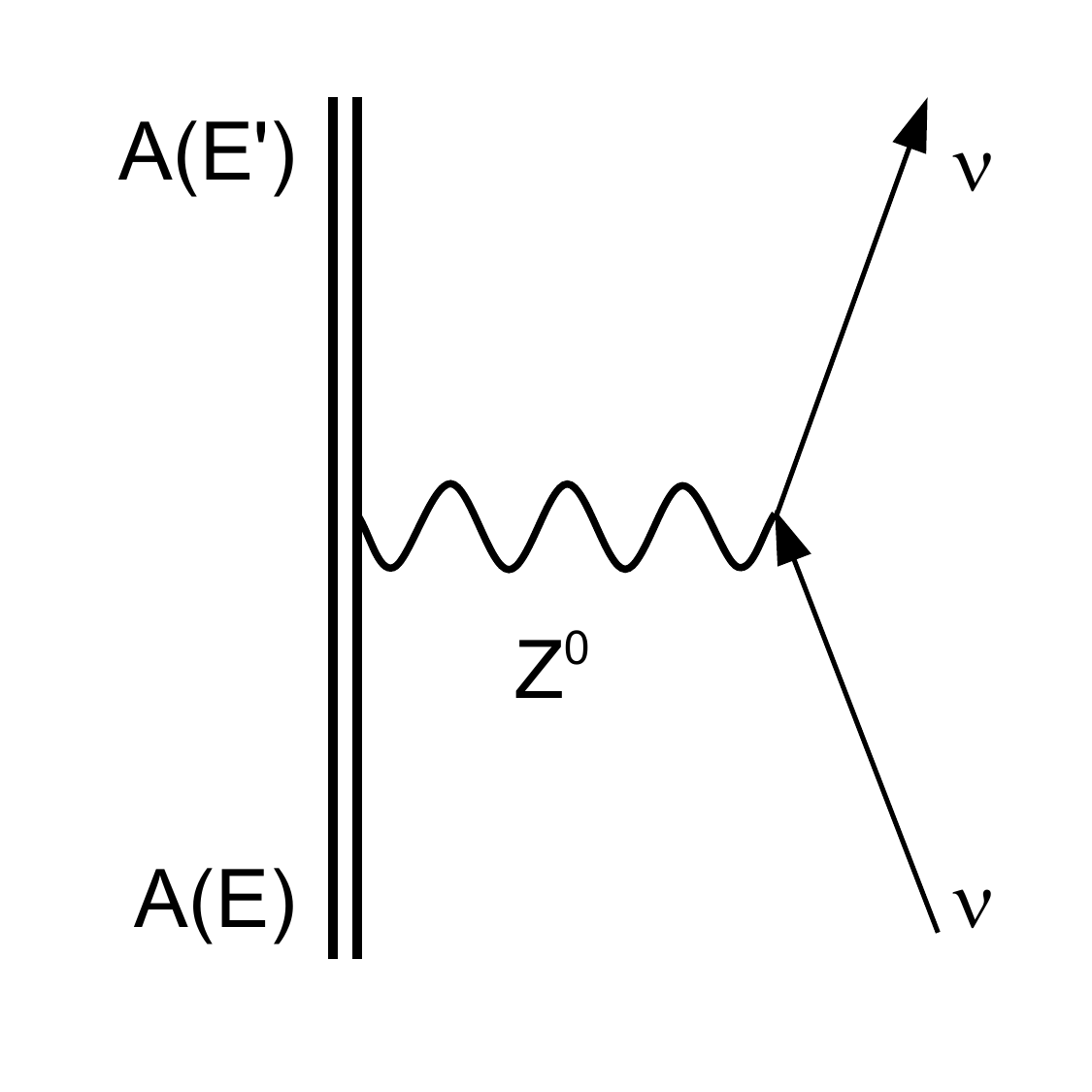}
\caption{Neutral current neutrino scattering from a nucleus of mass number A with initial excitation energy E and final excitation E'.}
\label{fig:feyn_nc_scatter}
\end{figure}

One final process that we will not discuss but which falls under the general purview of nuclear neutrinos is neutral current inelastic neutrino scattering on nuclei (figure \ref{fig:feyn_nc_scatter}) \cite{wh:1988,woosley-etal:1990,fm:1991}.  Scattering does not produce neutrinos, but it can alter the neutrino spectrum.  During the event, the nucleus can either gain internal energy from the neutrino in a subelastic scatter, or the nucleus can give up energy to the neutrino in a superelastic scatter.  The former will shift the neutrino spectrum down in energy, while the latter will shift it up.  Under most circumstances, there will be greater strength for a nuclear ``up-transition'' \cite{fm:1991}, meaning a subelastic scatter that lowers neutrino energy.  However, in supernova environments, there may be a sufficient population of excited nuclei to shift part of the neutrino spectrum up, lengthening the high energy tail of the spectrum, with possible implications for detection.

Sec. II details the calculation of the charged current process neutrino spectra and shows some results of high temperature shell model calculations.  In Sec. III we discuss neutral current nuclear de-excitation neutrino production and spectra, and in Sec \ref{sec:discussion}, we go over the results and their implications.

\section{Charged Current Process Neutrinos}
Both sd- and pf-shell nuclei will play a role in determining the pre-collapse neutrino and antineutrino energy spectra and corresponding fluxes.  Though we concentrate here on sd-shell nuclei, many of our conclusions on the role of nuclear excited states are also true for pf-shell nuclei.  We will speculate on nuclear structure issues for pf-shell nuclei in the section \ref{sec:discussion} discussion.

We carried out nuclear structure computations in the following manner.  We used the shell model code OXBASH \cite{oxbash} to compute energy levels and transition matrix elements of {\it sd}-shell nuclei.  Our model consisted of a closed $^{16}$O core, with the remaining nucleons unrestricted within the {\it sd} shell; the $1d_{5/2}$, $2s_{1/2}$, and $1d_{3/2}$ single particle states comprise the {\it sd} shell.  We employed the USDB Hamiltonian \cite{br:2006}.  Where feasible, we used experimentally determined nuclear state energies (for most of the nuclei we examined, these were the lowest 10 to 20 positive-parity states) and transition strengths (taken from published log(ft) values), and we otherwise relied on matrix elements calculated with the code.  We quenched the computed (non-experimental) squared transition matrix elements by a factor of 0.6.  We give rates and spectra in $s^{-1}baryon^{-1}$, and these are calculated as though the \emph{entire stellar core} were comprised of that material; that is, we show the emissivity of e.g. $^{21}$Ne as though the entire stellar core were $^{21}$Ne.

We first consider four charged current interactions: electron capture, positron emission, positron capture, and electron emission.  The former two produce electron flavored neutrinos via Gamow-Teller and Fermi isospin-lowering nuclear transitions (GT$^-$ and F$^-$), while the latter produce electron flavored anti-neutrinos via GT$^+$ and F$^+$ transitions.

\subsection{Charged current process calculation}
  To compute the rates, we follow the prescription of FFNI \cite{ffn:1980}.  The transition rate is given by

\begin{equation}
\lambda_{if}={\rm ln}2\frac{f_{if}(T,\rho Y_e)}{ft_{if}}
\end{equation}
where $ft_{if}$ is the relative half-life of the transition (this factor contains physical constants and the transition matrix elements), and $f_{if}(T,\rho Y_e)$ is the phase space factor.  We take

\begin{equation}
\frac{1}{ft_{if}}=\frac{1}{ft_{if}^{GT}}+\frac{1}{ft_{if}^F}=\frac{B_{if}^{GT\pm}}{10^{3.596}}+\frac{B_{if}^F}{10^{3.791}},
\end{equation}
which has units of $s^{-1}$, where
\begin{eqnarray}
B_{if}^{GT\pm}=\frac{\vert M_{if}^{GT\pm}\vert^2}{2J_i+1}=\frac{\vert \langle f\vert \sum\limits_k(\overrightarrow{\sigma}\tau^\pm)_k\vert i\rangle\vert^2}{2J_i+1} \\
B_{if}^{F\pm}=\frac{\vert M_{if}^{F\pm}\vert^2}{2J_i+1}=\frac{\vert \langle f\vert \sum\limits_k(\tau^\pm)_k\vert i\rangle\vert^2}{2J_i+1}.
\end{eqnarray}
Here, $\overrightarrow\sigma$ is the one-body spin operator, $\tau^\pm$ is the one-body isospin raising (upper sign) and lowering (lower sign) operator, $\vert i\rangle$ and $\vert f\rangle$ are the initial and final nuclear states, respectively, and the sums are over nucleons.

The phase space factor for decay processes is

\begin{equation}
f_{if}=\int_1^{-q}w^2(-q-w)^2G(Z,w)(1-f_{e^\pm}(w))dw
\end{equation}
and for capture processes is

\begin{equation}
f_{if}=\int_{max(1,q)}^{\infty}w^2(-q+w)^2G(Z,w)f_{e^\pm}(w)dw
\end{equation}
where w is the electron energy, q is the energy of the transition (daughter energy minus parent energy, including rest mass), G is the coulomb correction factor (described in detail in FFNI), and $f_{e^\pm}(w)$ is the electron (positron) distribution function; all energies are in units of electron mass.  Finally, these individual transition rates are summed over final states and thermally populated initial states.

Here, however, we are interested in neutrino spectra, rather than total rates; we obtain these by changing the phase space factor variable of integration from the electron energy to the neutrino energy, then keep only the kernel and do not integrate.  After changing variables, we interpret the kernel as the contribution to the transition rate per unit energy of the outgoing neutrino.  Since the spectrum is an explicit function of neutrino energy, we want a more convenient unit, so we factor $1/m_e$ out of all energies (yielding a total of four powers of $1/m_e$); this gives the freedom to choose any unit of energy.  To more readily compare nuclei, we divide by the mass number $A$.  Now, the units of the spectral density are neutrinos per second per baryon per electron mass; we divide the spectral density by one more power of $m_e$ to allow our choice of energy unit in the denominator.  As a consequence, we have

\begin{equation}
\begin{split}
S_{if}&(E_\nu) = \frac{{\rm ln}2}{A~m_e^5}\left(\frac{B_{if}^{GT\mp}}{10^{3.596}}+\frac{B_{if}^{F\mp}}{10^{3.791}}\right) \\
&E_\nu^2(-Q-E_\nu)^2G(Z,-Q-E_\nu)(1-f_{e^\pm}(-Q-E_\nu)) \\
&{\rm neutrinos}/s/{\rm baryon}
\end{split}
\end{equation}
for electron (lower signs) or positron (upper signs) decay and

\begin{equation}
\begin{split}
S_{if}&(E_\nu) = \frac{{\rm ln}2}{A~m_e^5}\left(\frac{B_{if}^{GT\pm}}{10^{3.596}}+\frac{B_{if}^{F\pm}}{10^{3.791}}\right) \\
&E_\nu^2(E_\nu+Q)^2G(Z,E_\nu+Q)f_{e^\pm}(E_\nu+Q) \\
&{\rm neutrinos}/s/{\rm baryon}
\end{split}
\end{equation}
for electron or positron capture, where $E_\nu$ is the neutrino energy and $Q$ is the nuclear transition energy.  We define $Q$ as the rest mass energy plus initial excitation energy of the daughter nucleus minus the rest mass energy plus final excitation energy of the parent nucleus: $Q=(M_d+E_{f})-(M_p+E_i)$.  We use MeV for all energies, so the units of the spectrum will be neutrinos per second per baryon per MeV.

We populate the initial states using the modification of the Brink hypothesis detailed in \cite{mfb:2014}, considering all states individually up to 12 MeV excitation, and assigning the remaining thermal statistical weight to the average of the next 50 or more higher states.  Finally, we sum over initial and final states.

\subsection{Charged current process spectra}
In this section, we choose nuclei, temperatures, and densities to facilitate comparison with P\&L.  That work made the excellent point that charged current processes could be the greatest source of high energy neutrinos.  The authors used the technique of Langanke et al \cite{lms:2001} to generate charged current neutrino spectra, whereby a single Q-value and transition strength for each nucleus are taken as parameters that are fit to published neutrino loss and energy loss rates (such as those of Fuller, Fowler \& Newman \cite{ffn:1980,ffn:1982a,ffn:1982b,ffn:1985}, Oda et al \cite{oda-etal:1994}, and Langanke \& Martinez-Pinedo \cite{lm:2001}).  That is, for each nucleus at a particular temperature and electron density (generally expressed as $\rho Y_e$), an effective Q-value and transition strength are chosen to reproduce the rates published for that nucleus in those conditions.  In contrast, we compute neutrino spectra by the method detailed in the previous section.

\begin{figure}
\includegraphics[scale=0.42]{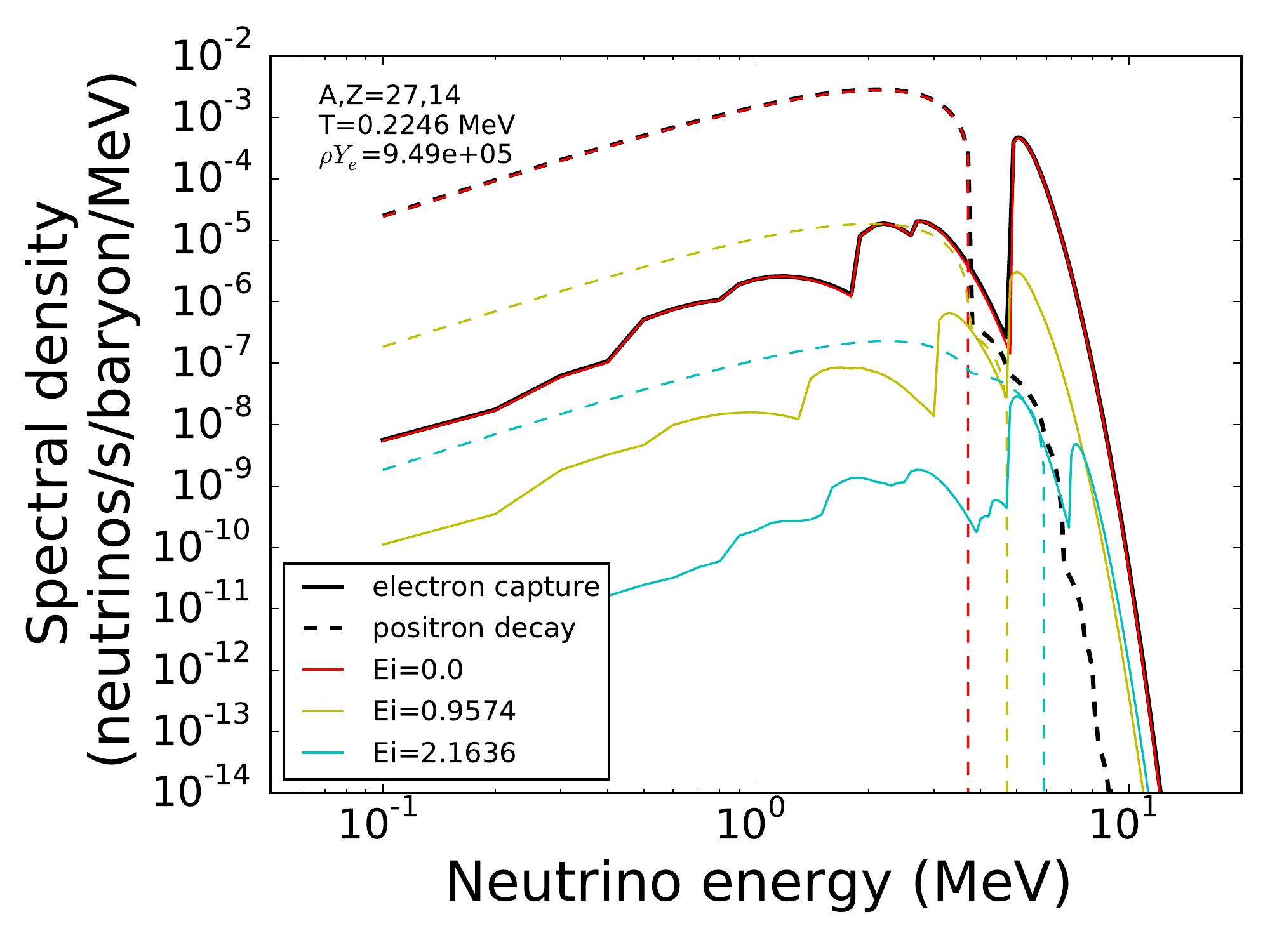}
\caption{(Color online) $^{27}$Si charged current process neutrino spectrum.  The black lines are totals for the nucleus, and the colored lines correspond to the indicated initial parent states.}
\label{fig:27si_cc}
\end{figure}

Figure \ref{fig:27si_cc} shows the neutrino energy spectrum from GT$^-$ and F$^-$ transitions from $^{27}$Si to $^{27}$Al at a temperature of $\sim 0.22$ MeV and $\rho Y_e\sim 9.5\times 10^{5}$ g/cm$^3$.  Solid lines are from electron capture, and dotted lines are from positron decay.  Black lines are totals for each process, while colored lines are the contributions from selected parent nucleus initial states. Transitions from the ground state dominate the spectrum almost everywhere, though naturally the positron decay spectrum at high neutrino energy comes from excited states.  Furthermore, $^{27}$Si has a proton excess of exactly 1, while $^{27}$Al has a neutron excess of exactly 1, so these two nuclei comprise a mirror system.  That is, from a structural perspective, these nuclei are identical up to a relabeling of protons and neutrons.  Therefore, each state in $^{27}$Si will have a superallowed (Fermi) transition to the corresponding state in $^{27}$Al, and in particular, the ground states are therefore connected.  Consequently, this transition has a tremendous amount of strength relative to other transitions, and it defines the shape of the spectrum: there is a single large peak from positron decay, and a single large peak from electron capture.  The smaller peaks in the electron capture channel arise from transitions to excited states in the daughter nucleus, but outside the narrow valley at $\sim 4$ MeV, they are buried under the positron decay peak.

\begin{figure}
\includegraphics[scale=0.42]{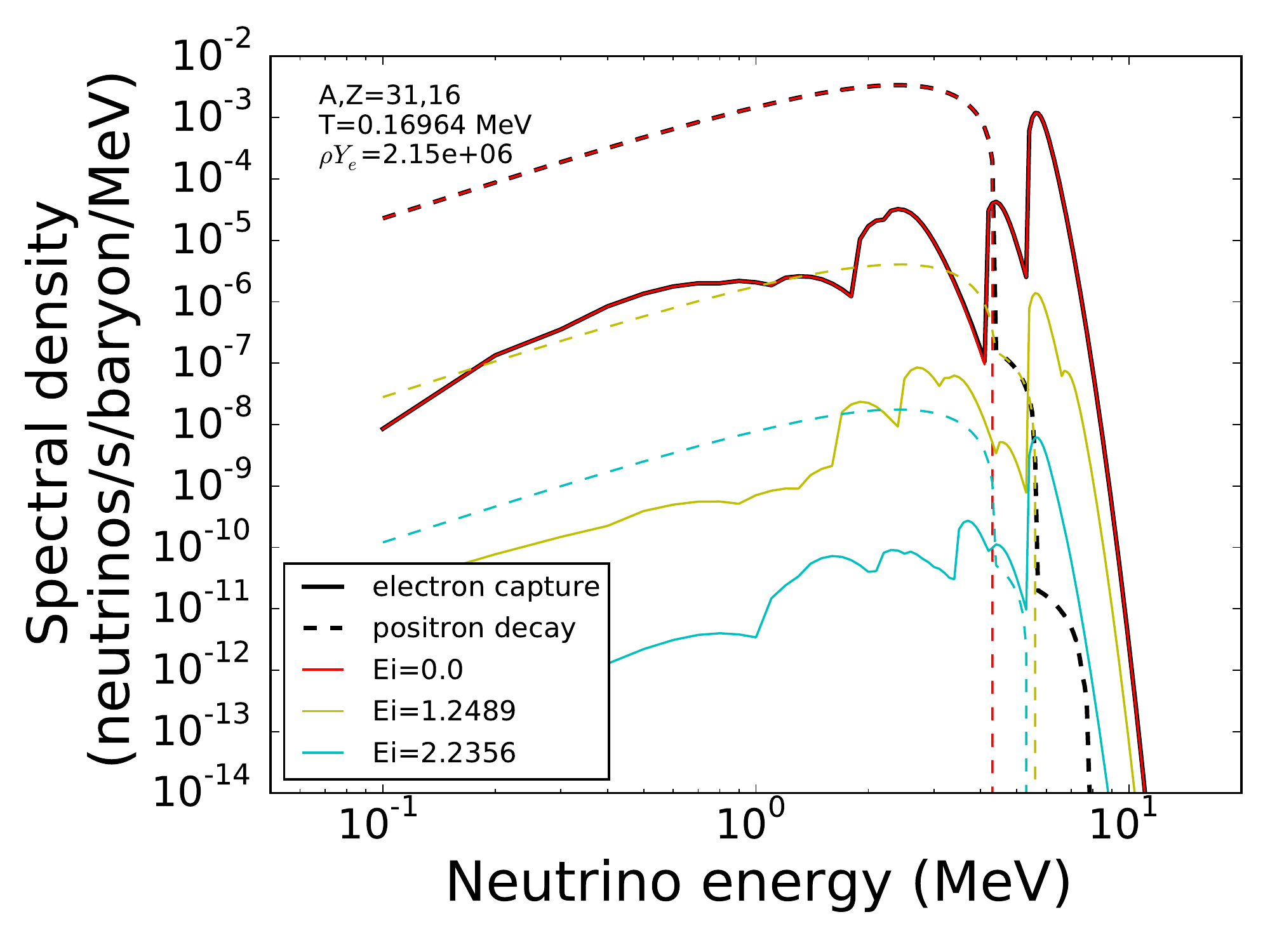}
\caption{(Color online) $^{31}$S charged current process neutrino spectrum.  The black lines are totals for the nucleus, and the colored lines correspond to the indicated initial parent states.}
\label{fig:31s_cc}
\end{figure}

Figure \ref{fig:31s_cc} (same line designations as in figure \ref{fig:27si_cc}) shows the neutrino spectrum from GT$^-$ and F$^-$ transitions from $^{31}$S to $^{31}$P at a temperature of $\sim 0.17$ MeV and $\rho Y_e\sim 2.2\times 10^{6}$ g/cm$^3$.  As with $^{27}$Si and $^{27}$Al, these are mirror nuclei, and as in figure \ref{fig:27si_cc}, the ground state-to-ground state transitions dominate the spectrum.

\begin{figure}
\includegraphics[scale=0.42]{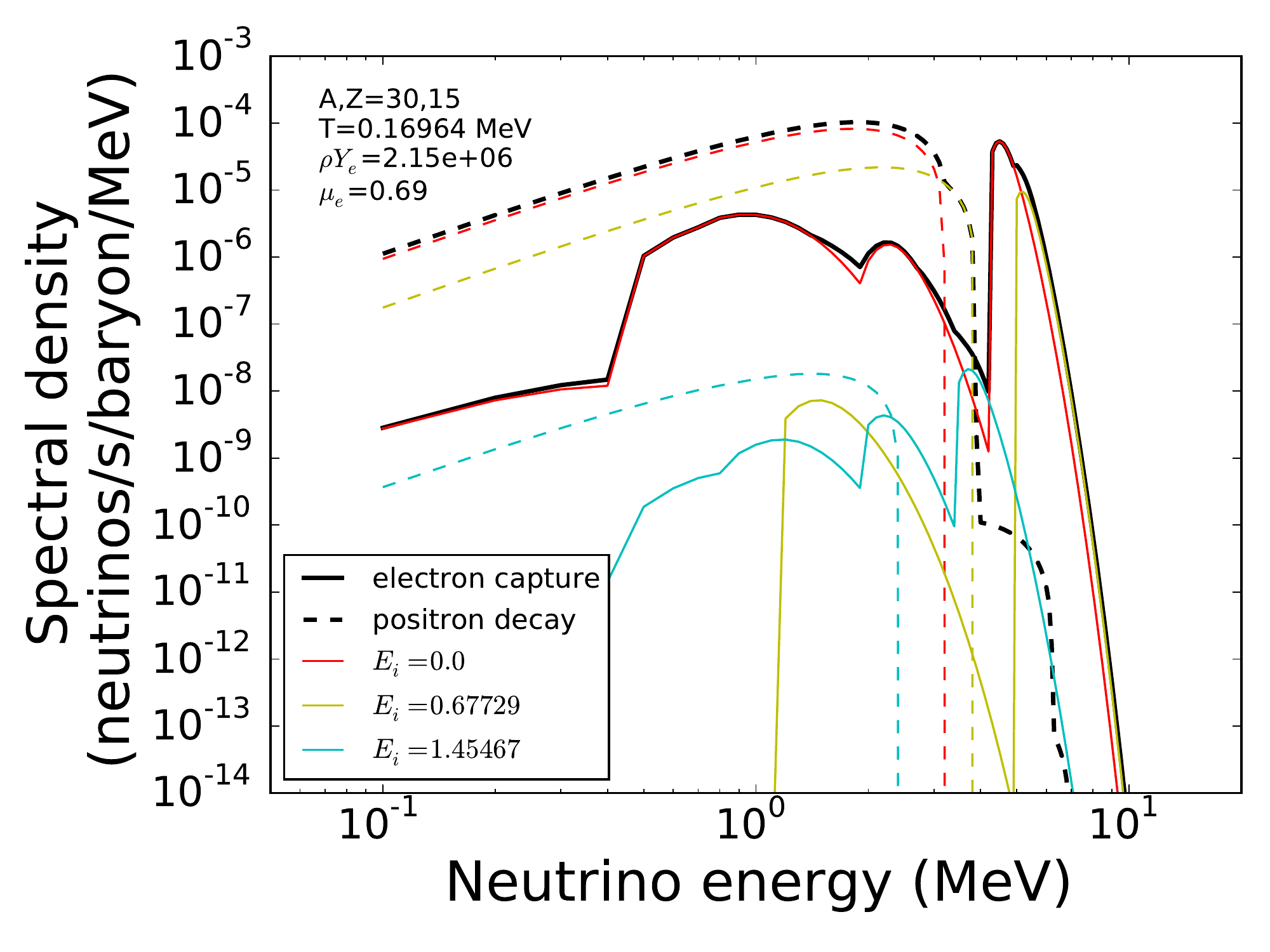}
\caption{(Color online) $^{30}$P charged current process neutrino spectrum.  The black lines are totals for the nucleus, and the colored lines correspond to the indicated initial parent states.}
\label{fig:30p_cc}
\end{figure}

Figure \ref{fig:30p_cc} (same line designations as in figure \ref{fig:27si_cc}) shows the neutrino spectrum from GT$^-$ and F$^-$ transitions from $^{30}$P to $^{30}$Si at a temperature of $\sim 0.17$ MeV and $\rho Y_e\sim 2.2\times 10^{6}$ g/cm$^3$.  This is not a mirror system, so the ground states are not connected by a Fermi transition.  Nevertheless, transitions from the ground state of $^{30}$P define the neutrino energy spectrum for energies $<3$ MeV and between 4 and 5 MeV.  However, the first excited state of $^{30}$P ($E_i=0.677$ MeV) is isospin $T=1$ and spin $J=0$, so it \emph{does} have a superallowed transition to the ground state of $^{30}$Si.  The strength of this superallowed transition causes it to dominate the neutrino spectrum from 3 to 3.9 MeV and above 5 MeV neutrino energy.  In the narrow band between the high-energy cutoff at 3.9 MeV for positron emission from the first excited state and the low energy cutoff at 4.2 MeV for electron capture on the ground state, the third excited state ($E_i=1.145$ MeV) produces most of the neutrinos.  Of course, the population of these excited states depends sensitively on temperature through the Boltzmann factors, so the conclusions may be different at lower temperature or higher temperature than we consider here.

\begin{figure}
\includegraphics[scale=0.42]{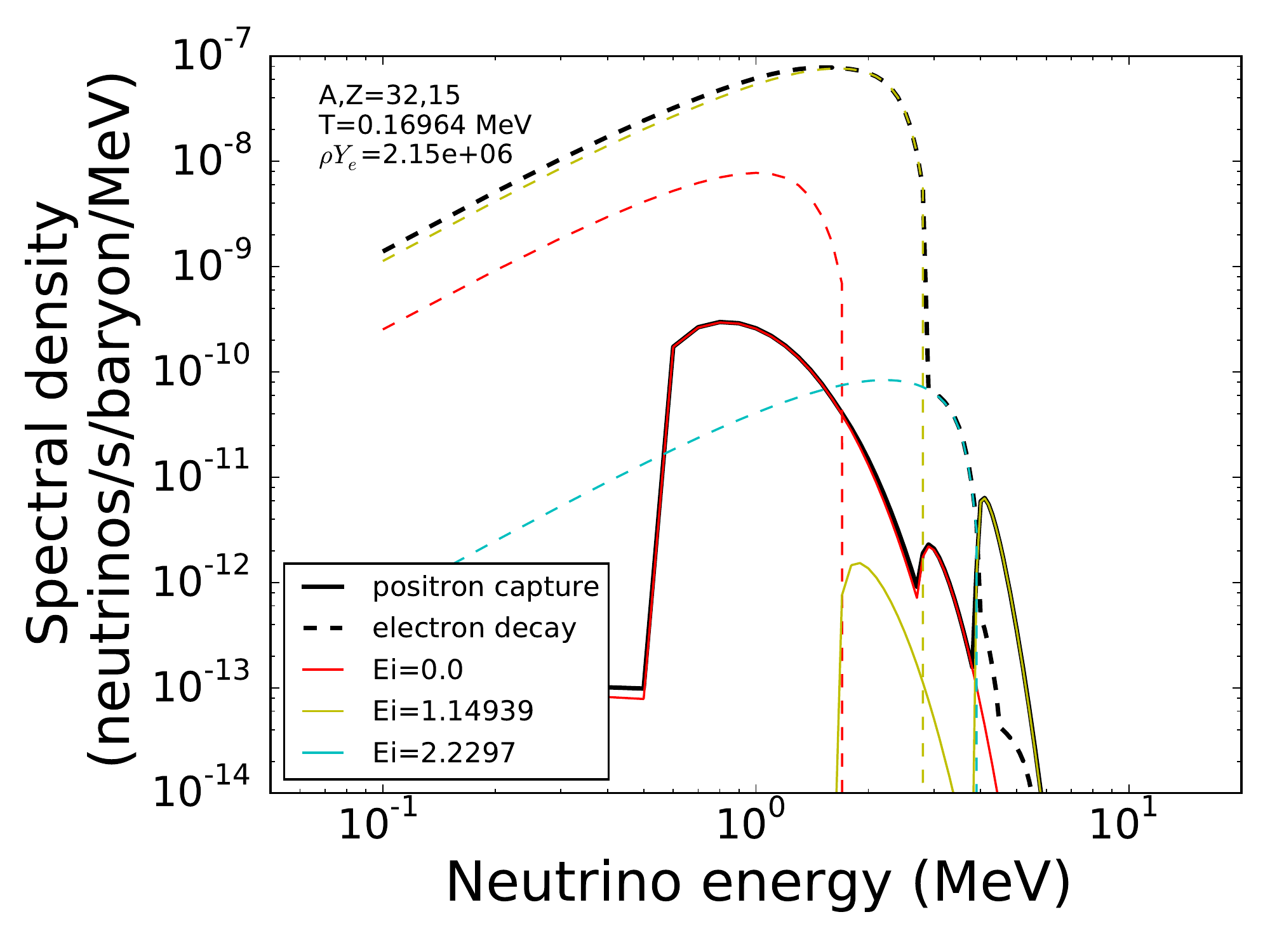}
\caption{(Color online) $^{32}$P charged current process anti-neutrino spectrum.  The black lines are totals for the nucleus, and the colored lines correspond to the indicated initial parent states.}
\label{fig:32p_cc}
\end{figure}

Figure \ref{fig:32p_cc} (same line designations as in figure \ref{fig:27si_cc}) shows the anti-neutrino spectrum from GT$^+$ and F$^+$ transitions from $^{32}$P to $^{32}$S at a temperature of $\sim 0.17$ MeV and $\rho Y_e\sim 2.2\times 10^{6}$ g/cm$^3$.  Here, while the ground state does dominate the total positron capture rate, it \emph{nowhere} dominates the spectrum.  Instead, transitions from the third excited state ($E_i=1.15$ MeV) produce most of the spectrum, with the band between 2.7 and 3.9 MeV coming from transitions from the eighth excited state ($E_i=2.23$ MeV).

The high-energy peak in the positron capture channel on the ground state of $^{32}$P arises from transitions to the ground state of $^{32}$S.  However, this transition has a rather low strength of $B_{GT}=0.50\times 10^{-4}$, while the transition to the first excited state of $^{32}$S has a very large strength of $B_{GT}=0.12$; consequently most of the positron capture neutrinos have low energy.  The third excited state ($E_i=1.15$ MeV), on the other hand, has a high strength transition to the ground state of $^{32}$S, with $B_{GT}=0.074$.  The high strength and large phase space factor overcome the Boltzmann factor relative to ground, resulting in this state being the principle source of electron decay neutrinos and high energy positron capture neutrinos.  The $^{32}$P $E_i=2.23$ state also has a fairly high strength transition to $^{32}$S ground ($B_{GT}=0.015$), so electron decay from this state fills the gap between the $E_i=1.15$ MeV electron decay and positron capture peaks.

\begin{figure}[here]
\includegraphics[scale=0.42]{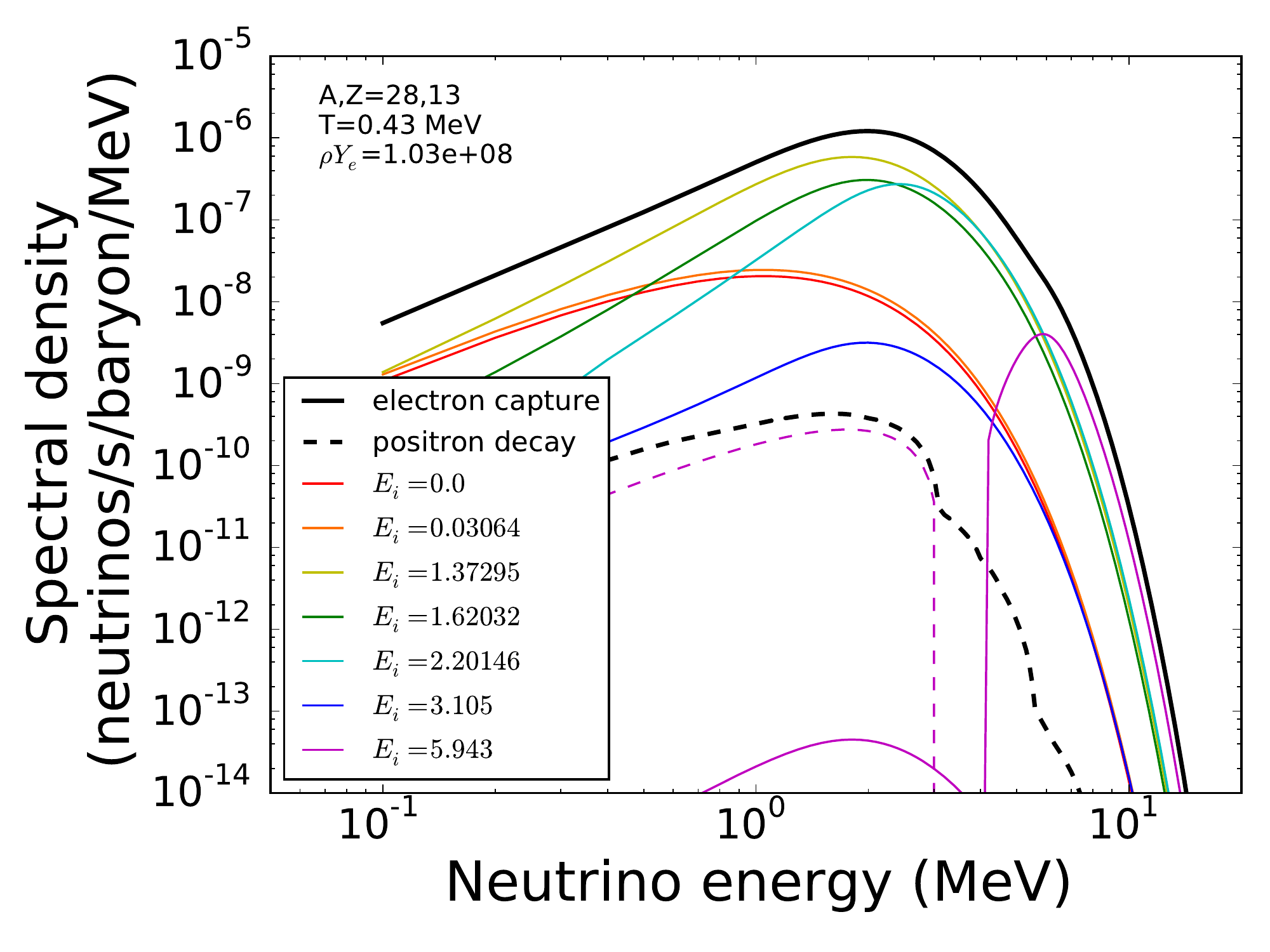}
\caption{(Color online) $^{28}$Al charged current process neutrino spectrum.  The black lines are totals for the nucleus, and the colored lines correspond to the indicated initial parent states.}
\label{fig:28al_cc}
\end{figure}

Figure \ref{fig:28al_cc} (same line designations as in figure \ref{fig:27si_cc}) shows the neutrino spectrum from GT$^-$ and F$^-$ transitions from $^{28}$Al to $^{28}$Mg at a temperature of $\sim 0.43$ MeV and $\rho Y_e\sim 1.0\times 10^{8}$ g/cm$^3$.  $^{28}$Al is lighter than the typical nucleus in these conditions, but it has close to the correct electron fraction and is therefore an interesting case.  The lowest four states of this nucleus have no allowed transitions to the ground state of $^{28}$Mg, and as a consequence, they are not significant contributors to the neutrino spectrum at any energy.  The 4th, 5th, and 8th excited states ($E_i=1.37$, 1.62, and 2.20 MeV, respectively) all have allowed transitions to ground and produce most of the neutrinos by electron capture.  The 15th excited state ($E_i=3.11$ MeV) has an allowed transition to $^{28}$Mg ground, but is hindered by a small thermal population factor and a fairly low transition strength ($B_{GT}=1.243\times 10^{-4}$).  The first isobaric analog state in $^{28}$Al occurs at $E=5.94$ MeV and is the principle source of positron decay neutrinos and very high energy electron capture neutrinos.

\section{Neutral Current De-excitation Neutrinos}
\subsection{Calculation of neutrino pair rates}
The de-excitation rate via neutrino pair production from an initial state $\vert i\rangle$ with energy $E_{i}$ to a final state $\vert f\rangle$ with energy $E_{f}$ is \cite{btz:1974}

\begin{equation}
\begin{aligned}
\lambda_{if} &\approx \frac{G^{2}_{F}g^{2}_{A}}{60\pi^{3}}\left(E_{i}-E_{f}\right)^{5}B^{GT3}_{if} \\
& \approx 1.71\times 10^{-4}s^{-1}\left(\frac{E_i-E_f}{MeV}\right)^{5}B^{GT3}_{if}.
\end{aligned}
\label{eq:de-excitation_rate}
\end{equation}
$G_F$ is the Fermi constant and $g_A$ is the axial vector coupling constant.  $B^{GT3}_{if}=\vert\langle f\vert\sum_k(\overrightarrow{\sigma} t_z)_k\vert i\rangle\vert^2/(2J_i+1)$ is the reduced squared matrix element for the transition; the sum is over nucleons, $\overrightarrow{\sigma}$ is the one-body spin operator, and $t_z$ is the z-component of the one-body isospin operator.

The energy loss rate is, of course, the de-excitation rate times the difference in initial and final state energy.  Including the thermal population probability of excited states and expressing the transition energy $\Delta E=\vert E_{f}-E_{i}\vert$ as a ratio to the ambient temperature, the energy loss rate per nucleus by de-excitation into neutrino pairs from state $\vert i\rangle$ to state $\vert f\rangle$ is

\begin{equation}
\begin{split}
\Lambda_{if}=1.71\times 10^{-4}\frac{\rm MeV}{s}\left(\frac{T}{\rm MeV}\right)^{6}\left(\frac{\Delta E}{T}\right)^{6} \\
\times B^{GT3}_{if}\frac{(2J_{i}+1)e^{-\left(\Delta E+E_{f}\right)/T}}{G(T)},
\end{split}
\label{eq:emission_rate}
\end{equation}
where $J_{i}$ is the spin of the initial state and $G(T)$ is the nuclear partition function at temperature $T$.

To guide the search for nuclei that might be important sources of neutrino pairs, we factor out of equation \ref{eq:emission_rate} the dimensionful factor $1.71\times 10^{-4}$ MeV/s, the factor $\left(T/{\rm MeV}\right)^{6}$, and those parts that depend explicitly on the characteristics of a particular nucleus, to whit, $B^{GT}_{if}(2J_{i}+1)/G(T)$, then apply an overall factor of $1/(6^{6}e^{-6})$ so that the peak ``normalized'' emission rate is 1.

\begin{equation}
\Lambda^{norm}_{if}=\frac{1}{6^{6}e^{-6}}\left(\frac{\Delta E}{T}\right)^{6}e^{-\Delta E/T}e^{-E_{f}/T}
\label{eq:norm_emission_rate}
\end{equation}

\begin{figure}
\includegraphics[scale=0.42]{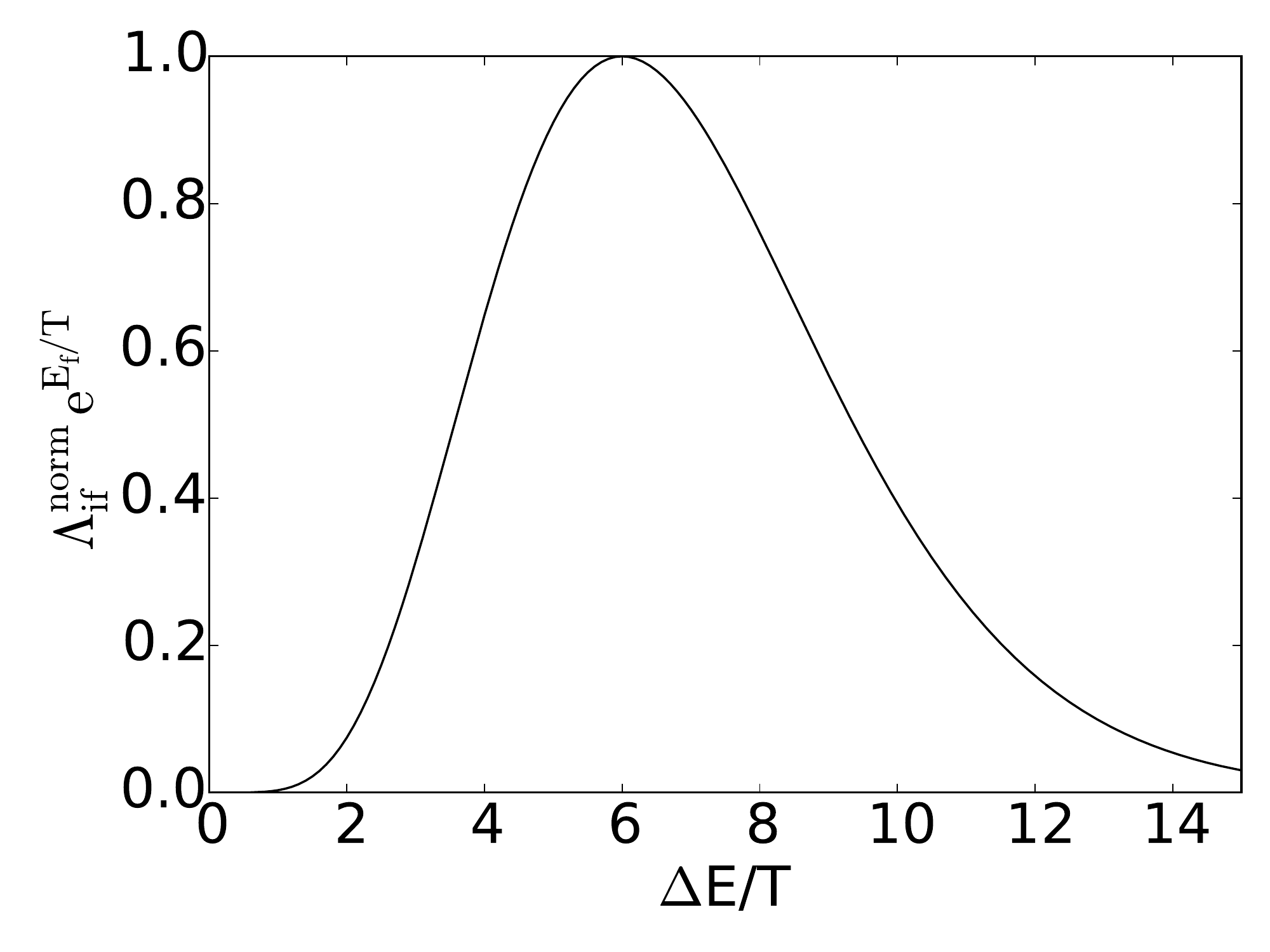}
\caption{Normalized energy emission rate in neutrino pairs via de-excitation to a final state with energy $E_{f}$ for a general nucleus as a function of the ratio of transition energy $\Delta E$ to temperature $T$.  This provides a qualitative guide to compare the emission rates of various nuclei at a given temperature.}
\label{fig:energy_effectiveness}
\end{figure}

From this expression, we see that which nuclei are effective at emitting energy in neutrino pairs depends on the ambient temperature: we seek nuclei with transitions and final states that are low enough in energy that the Boltzmann factor doesn't overly suppress the population, but balanced against that are the six powers of $\Delta E$ that favor higher excitations.  Figure \ref{fig:energy_effectiveness} shows the normalized emission rate (with the final state energy dependence factored out) as a function of $\Delta E/T$.

Equation \ref{eq:norm_emission_rate} shows that the peak in figure \ref{fig:energy_effectiveness} lies at $\Delta E/T=6$.  This means that we should look for nuclei that have transitions from excited states to low-lying (preferably ground) states with transition energies near $6T$.  The typical range of variation in $B^{GT}_{if}$ for transitions between low-lying states in sd-shell nuclei is about a factor of 10, so we can constrain our search to transition energies between about 3 and 10 times the temperature of the environment of interest.

Of course, equation \ref{eq:norm_emission_rate} applies to individual discrete transitions and does not take the density of states into account.  At low energy, states are sparse, so individual transitions tend to dominate the rate at low temperature.  As temperature increases, more and more states will fall on the high-energy slope of the peak in figure \ref{fig:energy_effectiveness}, reducing the importance of individual transitions and increasing the most effective energy.  As a consequence, at high temperature, the rates can be dominated by the density of states and the overall weak strength energy distribution, and we needn't be concerned about the detailed energy level structure in seeking important nuclei \cite{mfb:2014}.

\subsection{Energy loss rates}
\begin{figure}
\includegraphics[scale=.42]{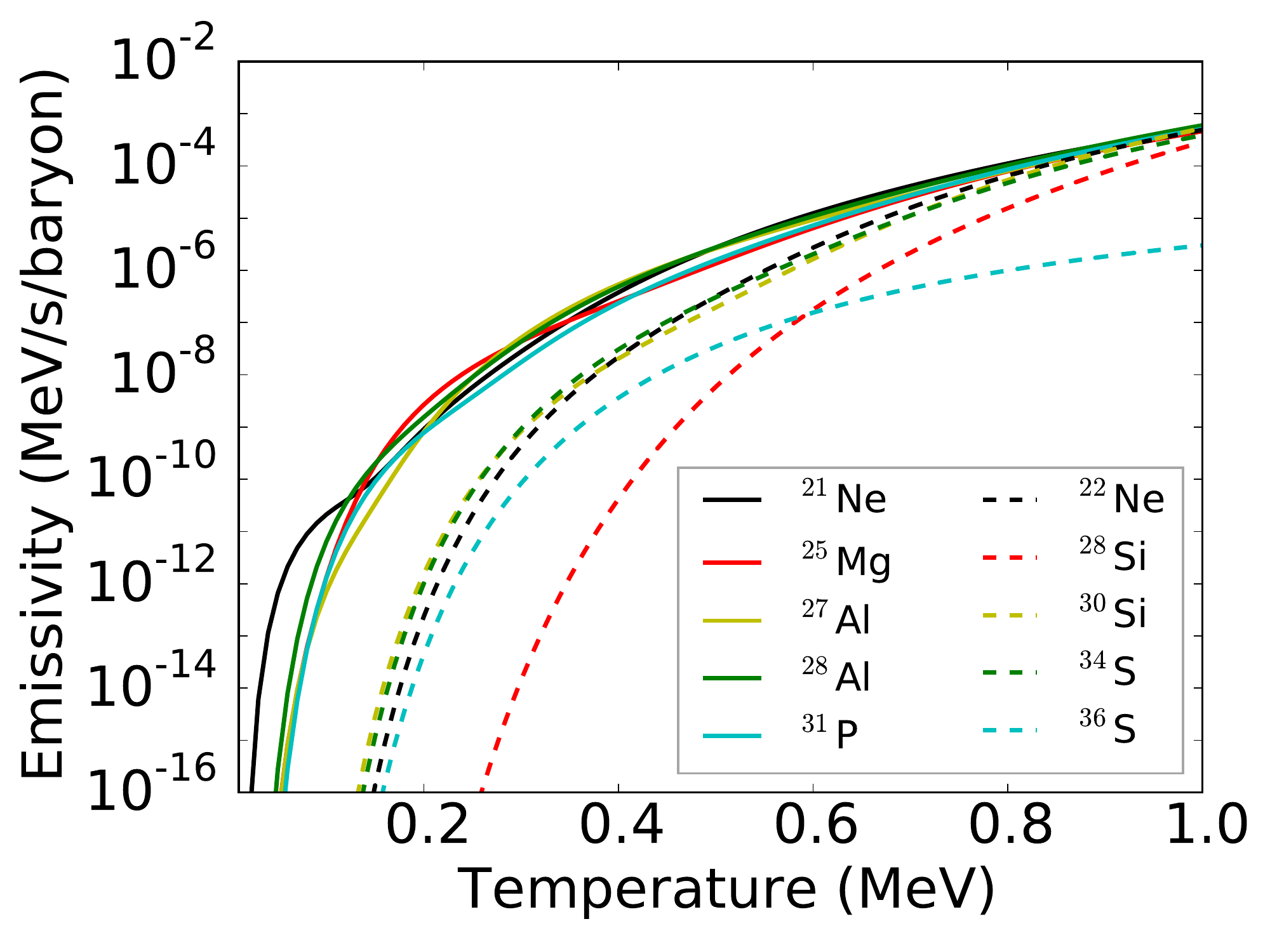}
\caption{Neutrino emissivities for a variety of sd-shell nuclei.  The odd nuclei stay tightly grouped over the entire range of temperature.  At low temperature, $^{28}$Si has low emissivity due to a lack of low-lying states.  At high temperature, $^{36}$S is low due to model space restriction.}
\label{fig:nuclear_rates}
\end{figure}

Figure \ref{fig:nuclear_rates} shows the energy loss rate via neutral current de-excitation of a variety of {\it sd}-shell nuclei.  The rates are computed by summing equation \ref{eq:emission_rate} over initial and final states.  Following reference \cite{mfb:2014}, we considered each state individually up to a cutoff energy, and the remaining statistical weight is carried by a single average high-energy state computed from a sample of states (50 or more) above the cutoff.  For each nucleus, we chose a cutoff of 10, 12, or 15 MeV according to how many states in that nucleus we had computed transition matrix elements for.  Included in the figure are a selection of odd-even nuclei (nuclei with an even number of protons and an odd number of neutrons or vice versa), the odd-odd nucleus $^{28}$Al, the four stable even-even {\it sd}-shell nuclei with relatively low-lying ($E<6$ MeV) $J^\pi=1^+$ states (which have allowed transitions to ground), and the tightly bound, difficult-to-excite nucleus $^{28}$Si.

Over the entire range of temperature, the odd nuclei remain tightly clustered in one group, and the even-even nuclei with low-lying $1^+$ states comprise a second group ($^{36}$S strays away from this group at high temperature due to model space restriction).  $^{28}$Si falls well below both groups at modest temperatures because it has very few allowed transitions between low-lying states.  At high temperature, all of the nuclei converge into a single group, indicating that the behavior at high temperature is independent of the specific nucleus.

\begin{figure}
\includegraphics[scale=.42]{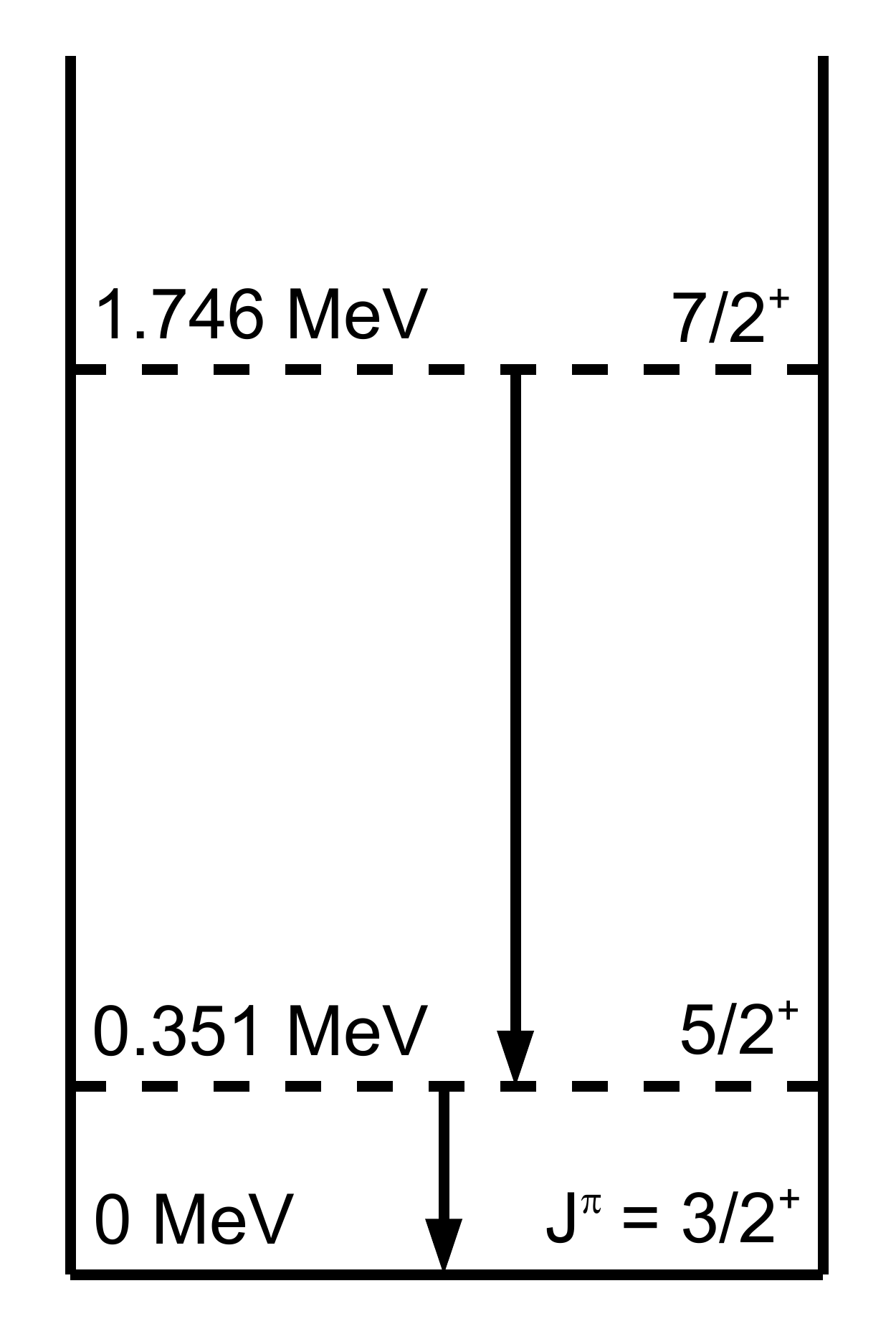}
\caption{There are two allowed de-excitations from low-lying excited states in $^{21}$Ne.  Figure \ref{fig:nuclear_rates} shows that each of these transitions becomes important at different temperatures.}
\label{fig:21ne_cartoon}
\end{figure}

The neutrino and anti-neutrino emissivity of $^{21}$Ne has two prominent ledges as temperature increases.  The first occurs at very low temperature ($<0.1$ MeV), causing $^{21}$Ne to dominate the other odd nuclei.  We can understand this behavior by examining the low-lying energy level structure of $^{21}$Ne.  Figure \ref{fig:21ne_cartoon} shows the three lowest-energy states.  The first rise is due to the exceptionally low-energy first excited state, which is substantially lower than the first excited state of each other nucleus in the figure (except for $^{28}$Al, which has a nearly-degenerate ground state).  Referring to figure \ref{fig:energy_effectiveness}, the transition from the first excited state to ground reaches peak relative effectiveness at $T\sim 0.06$ MeV, but the next allowed transition does not become effective until $T\sim 0.2$ MeV.  This illustrates the importance of individual transitions at low temperatures.

\begin{figure}
\includegraphics[scale=.42]{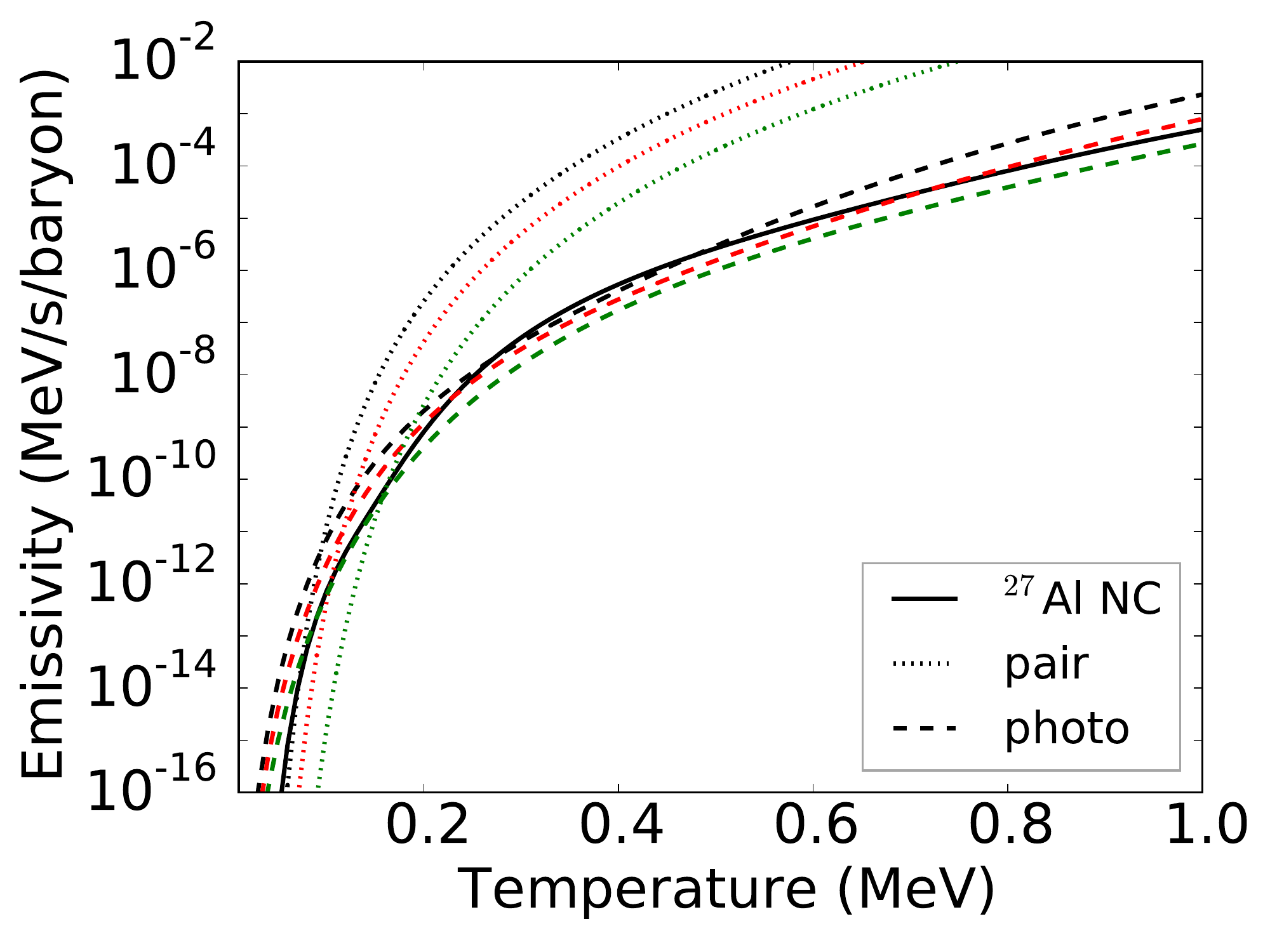}
\includegraphics[scale=.42]{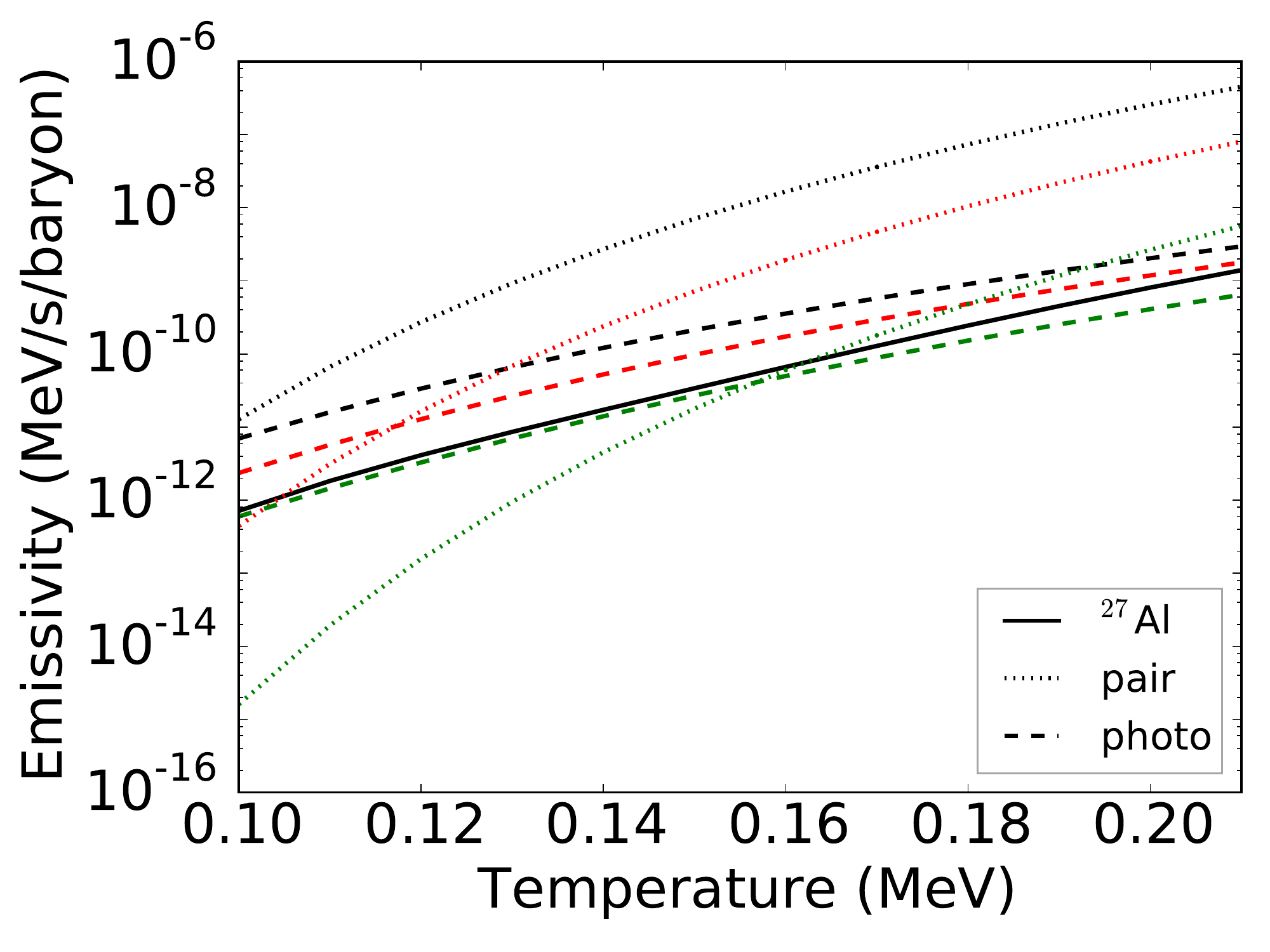}
\caption{Energy loss rates from neutral current de-excitation of $^{27}$Al and other major neutrino processes.  For electron-positron pair annihilation and the photo process, the black (upper) lines are for $\rho =10^7$ g/cm$^3$, the red (middle) lines are for $3\times 10^7$ g/cm$^3$, and the green (lower) lines are for $10^8$ g/cm$^3$.  The lower panel emphasizes the temperature range appropriate for core O-Ne-Mg burning.}
\label{fig:nc_vs_others}
\end{figure}

We also compare the energy loss rates from neutral current de-excitation against the other dominant sources of neutrino emission.  Over the temperature and density range relevant to core O-Ne-Mg burning and Si burning, the other two dominant sources of neutrino pairs are electron-positron pair annihilation and the photo process \cite{itoh-etal:1996}.  Figure \ref{fig:nc_vs_others} shows the emissivities of $^{27}$Al (chosen to represent the odd-nuclei bundle), pair annihilation, and the photo process.  The rates for the latter two processes are sensitive to density (both decrease with increasing density), so we include the rates for $\rho=10^{7}$ (black, upper), $3\times 10^{7}$ (red, middle), and $10^{8}$ (green, lower) g/cm$^3$.  The bottom panel is a zoom-in on the top panel, emphasizing the temperature range relevant for O-Ne-Mg burning.  From these plots, we see that neutral current de-excitation is likely never the dominant source of energy loss via neutrino pairs, but it may nevertheless be a significant contributor.

\begin{figure}[here]
\includegraphics[scale=.42]{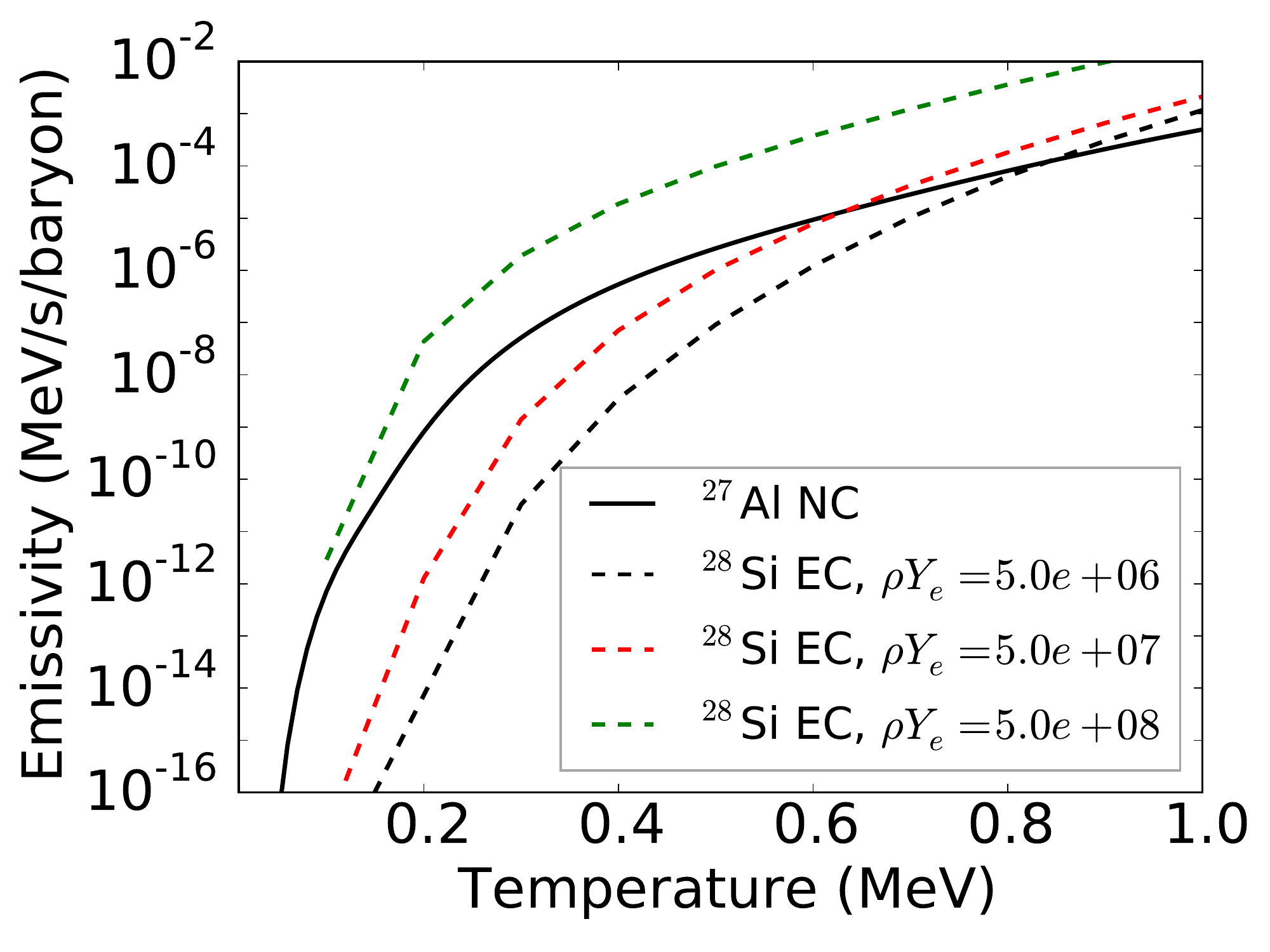}
\caption{Energy loss rates from neutral current de-excitation of $^{27}$Al and electron capture on $^{28}$Si.  The product of density and electron fraction $\rho Y_e$ is in g/cm$^3$.}
\label{fig:nc_vs_ec}
\end{figure}

Electron capture is the final major source of energy loss in highly evolved stellar cores.  In figure \ref{fig:nc_vs_ec}, we compare the emissivity of $^{27}$Al with the energy loss rate from electron capture on $^{28}$Si.  We computed the electron capture rate using the prescription of reference \cite{mfb:2014} with a cutoff of 15 MeV.  We include $\rho Y_e=5\times 10^6$ (black, lower), $5\times 10^7$ (red, middle), and $5\times 10^8$ (green, upper) g/cm$^3$.  The odd-nucleus neutral current rate dominates electron capture on $^{28}$Si until very late in silicon burning when the core is near collapse.  This comparison is somewhat unfair, however, as $^{28}$Si is an even-even nucleus, and odd nuclei tend to have a higher density of states and more allowed transitions.  We should be careful, therefore, to not draw broad conclusions from this one comparison and use it only as a guide for further exploration; to accurately compare the two processes, we must know the abundances and relative production rates of many nuclei.

\subsection{Neutral current spectra}
Although neutral current de-excitation of nuclei is probably not a major source of energy loss in O-Ne-Mg burning and Si burning, the energies of the pairs can be much greater than the typically thermal energies of pairs produced by other processes.  The spectral density of neutrinos from de-excitation from initial nuclear state $\vert i\rangle$ to final state $\vert f\rangle$ is computed similarly to the spectrum from charged current processes.  In this case, the kernel of the phase space integral that yielded equation \ref{eq:de-excitation_rate} is simply $E_\nu^2(-Q-E_\nu)^2$, giving a spectral density of

\begin{equation}
\begin{split}
S_{if}(E_\nu)=5.134\times 10^{-3}\frac{B^{GT^3}_{if}}{A^{-1}}\left( \frac{E_\nu}{MeV}\right)^2\left( \frac{-Q-E_\nu}{MeV}\right)^2\\
{\rm neutrinos/s/baryon/MeV}
\end{split}
\label{eq:nc_spect}
\end{equation}

We sum equation \ref{eq:nc_spect} over final states and thermally populated initial states as before, producing a complete spectrum for the nucleus.

\begin{figure}
\includegraphics[scale=.42]{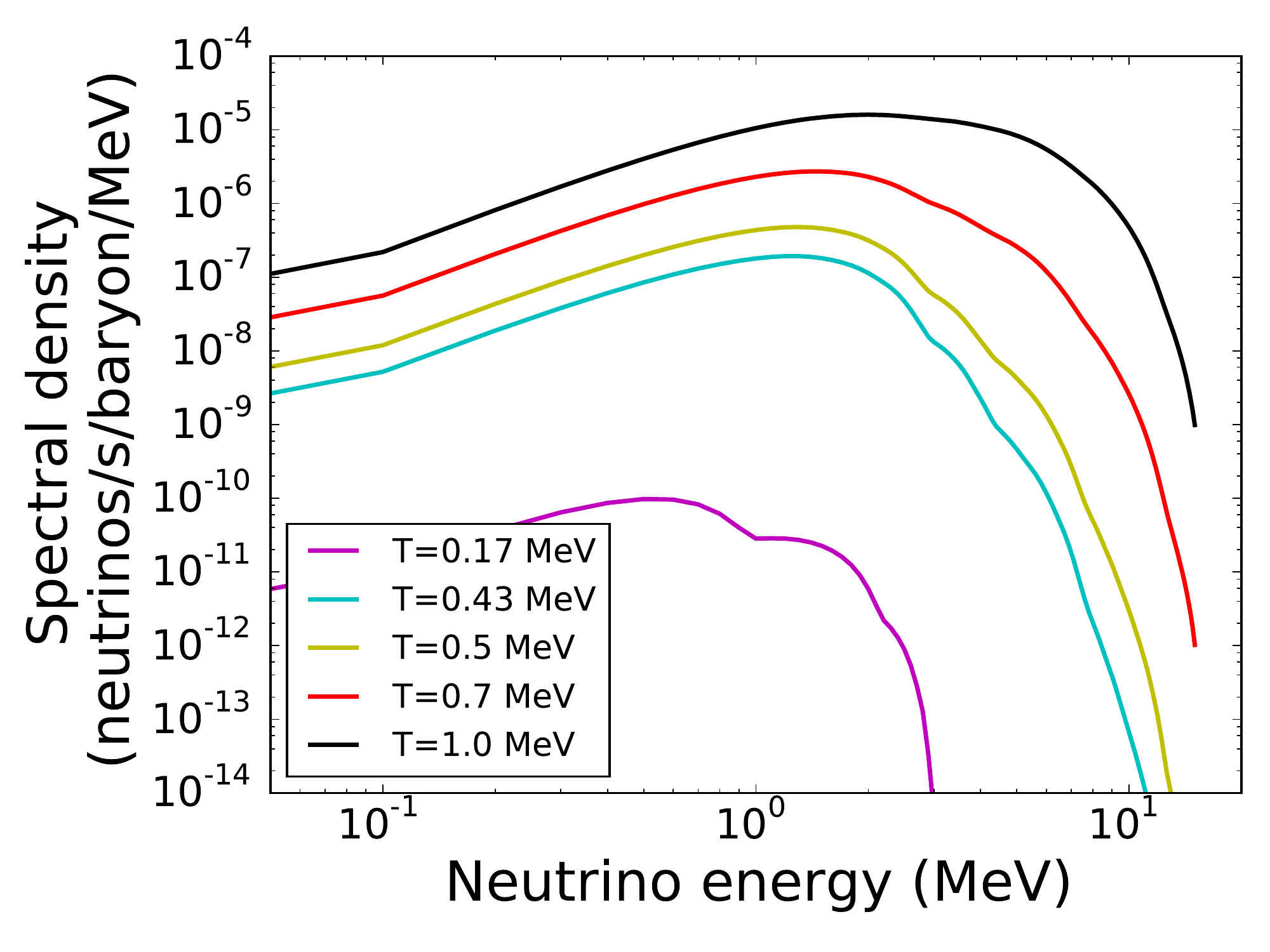}
\includegraphics[scale=.42]{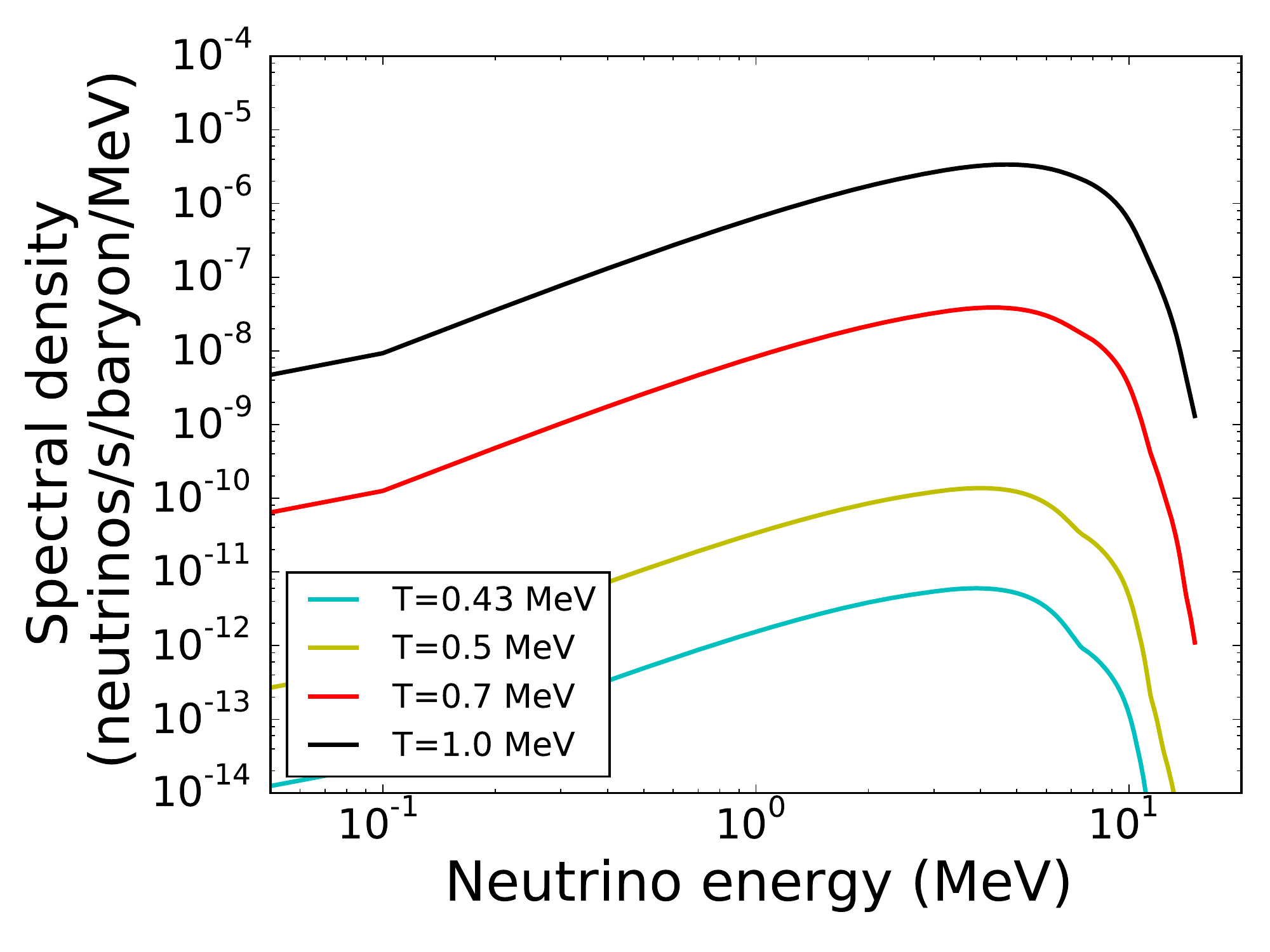}
\caption{$^{27}$Al (top) and $^{28}$Si (bottom) neutral current neutrino spectra.  Anti-neutrino spectra are identical.  Notably, the spectra above 10 MeV are similar.}
\label{fig:nc_spectra}
\end{figure}

Figure \ref{fig:nc_spectra} shows the neutrino spectra for $^{27}$Al (top panel) and $^{28}$Si (bottom panel) at a selection of temperatures relevant to late stellar evolution; by symmetry, the anti-neutrino spectra are identical.  We calculated these spectra using a further modification of the cutoff method detailed above, with the difference being that selections of states above the cutoff are grouped into energy bins which we average over; this avoids overpopulating the very high energy tails of the spectra.

Remarkably, above 10 MeV, the spectra of both nuclei are nearly identical.  That such different nuclei produce similar high energy neutrino spectra, coupled with the convergence of emissivities at high temperature in figure \ref{fig:nuclear_rates}, suggests that all {\it sd}-shell nuclei will produce similar results.  We included in figure \ref{fig:nc_spectra} very high temperatures relevant at the onset of and during core collapse.  An interesting feature of de-excitation pairs is that their rates and spectra are \emph{entirely independent} of the electron density, so that in a highly evolved pre-collapse and early collapse core, this might be a dominant source of high energy neutrinos.

\section{Discussion}
\label{sec:discussion}
Detecting neutrinos from highly evolved pre-collapse stars could give key insights into stellar evolution.  This is an exciting prospect.

P\&L astutely point out the importance of nuclear neutrinos in understanding late stellar neutrino spectra, and we build on that by examining the effects of nuclear structure.  To that end, we draw specific attention to $^{32}$P, shown in figure \ref{fig:32p_cc} of this work and figure 3 of P\&L \cite{pl:2015}.  In P\&L figure 3, there is a small bump in the anti-neutrino spectrum at $\sim 4$ MeV that the authors say is due positron capture on $^{32}$P.  At that point in P\&L's simulation, the mass fraction of $^{32}$P is $\sim 10^{-4}$ (personal communication).  Using the mass fraction and the density of the core, we convert the P\&L y-axis and find that the height of the P\&L $^{32}$P 4 MeV anti-neutrino peak is $\sim 8.5\times 10^{-10}$ neutrinos/second/baryon/MeV.  This corresponds roughly with the height of the $\sim 1$ MeV positron capture neutrino peak in our figure \ref{fig:32p_cc}.  By design, the single Q-value technique will give the correct total neutrino output with the correct average energy, but in this case, the energetics of the positron capture neutrinos are incorrect.  In this particular case, the published rates are dominated by electron emission from the first excited parent state, but most captures occur between the parent ground state and the first excited state of the daughter, pushing the positron capture neutrino energy down; this results in erroneous conclusions from the single Q-value method.  Similarly, the single Q-value method fails to capture the significant contribution of 3-4 MeV anti-neutrinos from the $E_i=2.23$ MeV state.

Finally, comparing figure \ref{fig:nc_spectra} in this work with the final plot in figure 4 of P\&L indicates that in late silicon burning, neutral current de-excitation may be a leading source of anti-neutrinos with energies greater than 10 MeV.  This is contradicted by the $^{28}$Al spectrum in this work's figure \ref{fig:28al_cc}, however, so we must be circumspect in drawing conclusions.  Nevertheless, it is clear that the production rates of $>10$ MeV neutrinos increase dramatically with temperature (due to the exponential dependence of the Boltzmann factor for excited states), and these rates are entirely unaffected by density and the associated baggage (such as electron blocking).  This implies that in this neutrino energy range, the effectiveness of this process relative to charged current processes is highly sensitive to the ambient temperature, and no solid conclusions can be drawn until realistic nuclear neutrino spectra are included in a simulation.  At higher temperatures--approaching the onset of collapse--neutral current de-excitation may be the dominant source of $>10$ MeV neutrinos.

\begin{figure}
\includegraphics[scale=.42]{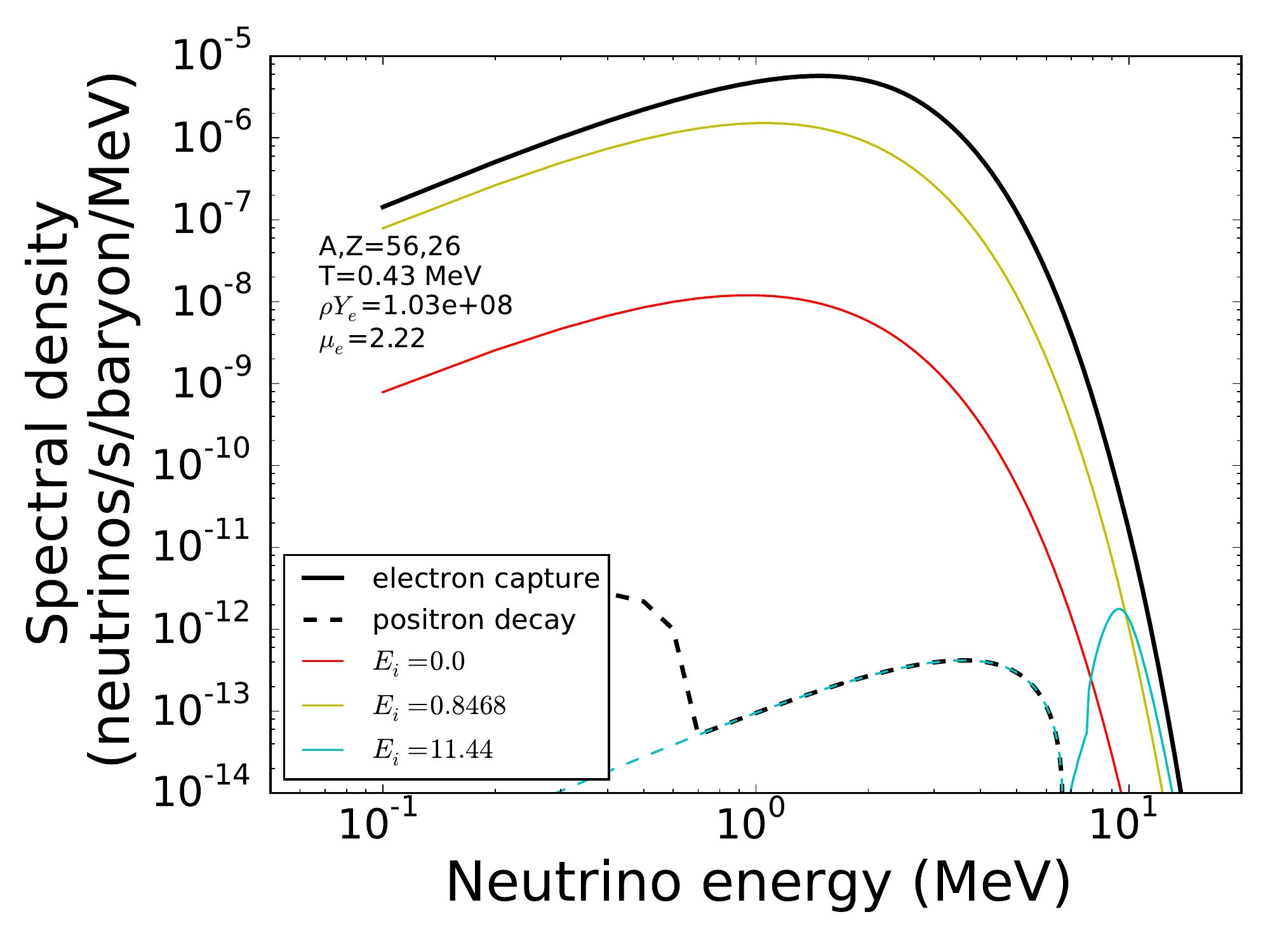}
\includegraphics[scale=.42]{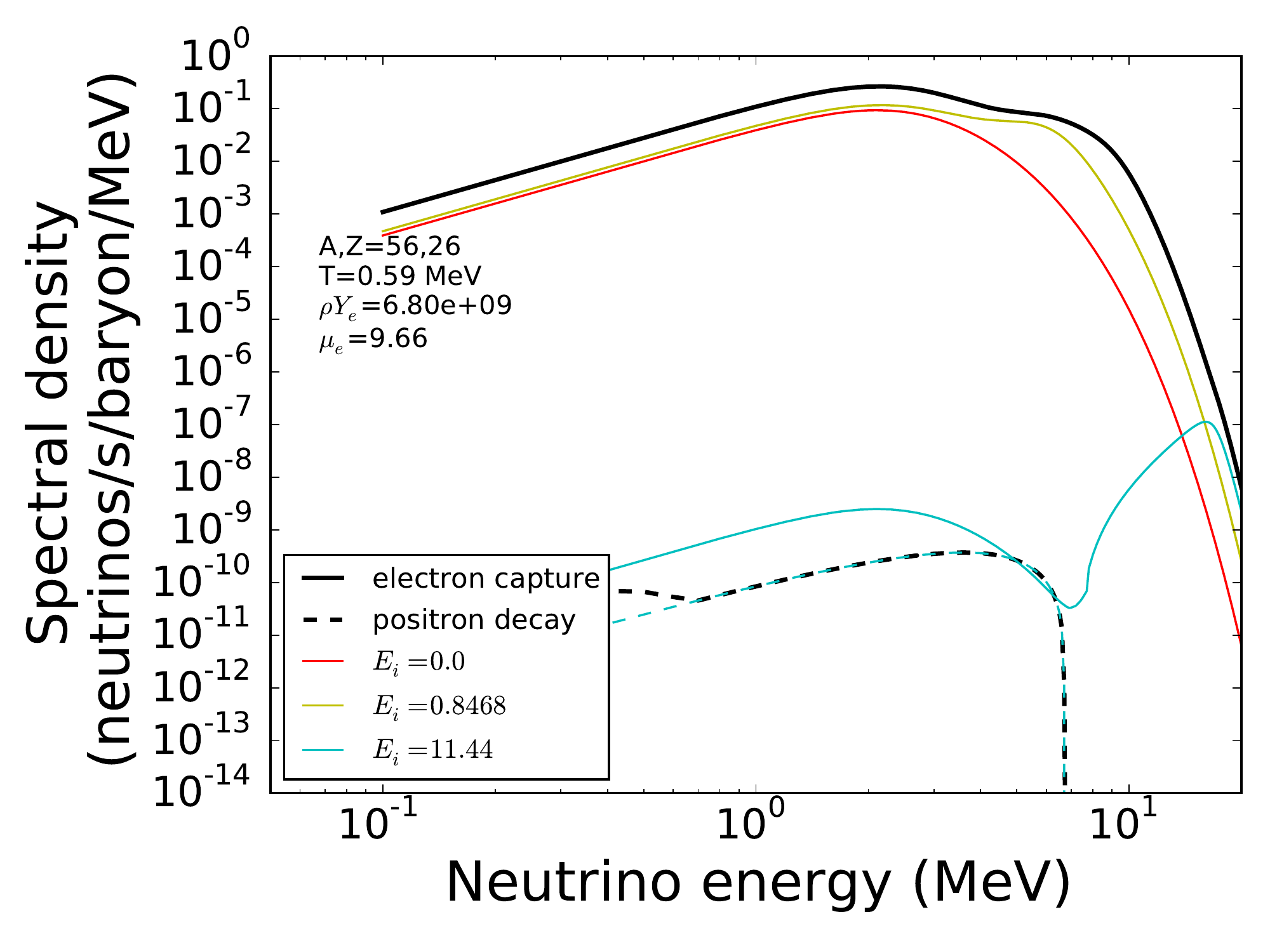}
\caption{$^{56}$Fe charged current process neutrino spectra computed from the FFN prescription.  The $E_i=11.44$ MeV line corresponds to the isobaric analog of the $^{56}$Mn ground state.  In the upper panel, the electron chemical potential is less than the Gamow-Teller resonance energy, while in the lower panel, it is greater than the GT resonance energy.  Because of this, the peak in the lower panel in more than 4 orders of magnitude greater, despite the comparatively small increase in temperature.}
\label{fig:56fe_cc}
\end{figure}

During core collapse, the electron chemical potential ($\mu_e$) climbs as density increases, with the consequence that the average energy of a captured electron is very high.  When $\mu_e$ reaches the energy of the Gamow-Teller resonance of a typical nucleus, the capture rate takes off, producing neutrinos prodigiously.  Following precisely the method of FFN, we computed electron capture and positron decay neutrino spectra for $^{56}$Fe.  Using the FFN prescription, $^{56}$Fe has a GT resonance at $\sim8$ MeV.  Figure \ref{fig:56fe_cc} (same line designations as in figure \ref{fig:27si_cc}) shows the spectra for two points leading up to and during collapse.  The upper panel has $\mu_e=2.22$ MeV (less than the resonance), and the lower has $\mu_e=9.66$ MeV (greater than the resonance).  The increase in temperature is not large, but bringing $\mu_e$ above the resonance energy increases the peak in the spectrum by more than 4 orders of magnitude.

\begin{figure}[here]
\includegraphics[scale=.42]{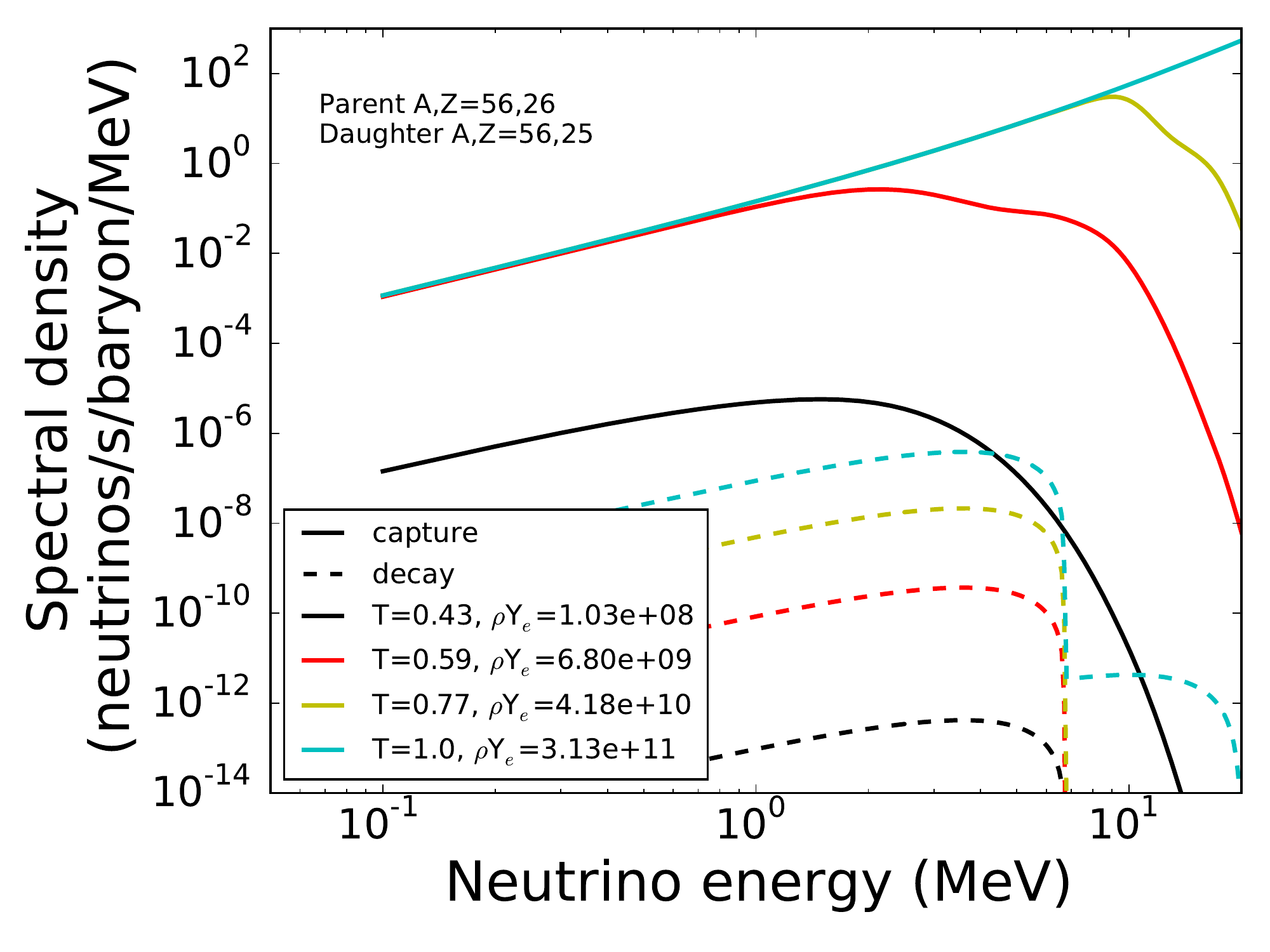}
\caption{$^{56}$Fe charged current process neutrino spectra computed from the FFN prescription at various points during collapse.  The enormous jump in neutrino production between the lowest two temperatures is due to the chemical potential in the higher temperature point being greater than the Gamow-Teller resonance energy.}
\label{fig:56fe_cc_multi}
\end{figure}

Figure \ref{fig:56fe_cc_multi} shows the $^{56}$Fe electron neutrino energy spectra computed using the FFN prescription at several points during collapse.  The solid lines are for electron capture, the dotted lines are for positron decay, and the colors correspond to different temperature and density conditions.  The results of figures \ref{fig:56fe_cc} and \ref{fig:56fe_cc_multi} are qualitative (strength is unquenched, delta function resonance, etc.) but indicate that at high $\mu_e$, the distribution of the bulk of the strength dominates the effects of precise structure.
  
Given the obvious importance of nuclear contributions to neutrinos with detectable energies, we will move forward in generating tabulated nuclear neutrino energy spectra in the same vein as the neutrino production and energy loss rates of earlier works.

\section{Acknowledgments}
We thank Kelly Patton for her insight and helpful discussions and Gang Guo for his assistance in computing the pair-annihilation and photo process energy loss rates.  We also thank B. Alex Brown for discussions and his assistance in carrying out and interpreting shell model computations.  We owe gratitude to Yang Sun and his group at SJTU for support and numerous discussions.  This work was supported in part by the National Natural Science Foundation of China (Nos. 11575112, 11135005) and by the 973 Program of China (No. 2013CB834401) at SJTU and NSF Grant No. PHY-1307372 at UCSD.

\bibliography{../references}

\end{document}